\documentclass[5p,times]{elsarticle}
\usepackage{lineno,hyperref}
\modulolinenumbers[5]

\journal{Computer Science Review}

\usepackage{colortbl}
  
\definecolor{White}{gray}{0.995}
\usepackage[ruled]{algorithm2e}
\usepackage[normalem]{ulem}
\usepackage{booktabs}
\usepackage{subcaption}
\usepackage{tabularx}
\usepackage{enumerate}
\usepackage{multirow}
\usepackage{amsmath}
\usepackage{lscape}
\usepackage{array}
\usepackage{breqn}
\usepackage{graphicx}
\usepackage{amssymb}
\usepackage{pifont}
\usepackage{csquotes}
\usepackage{tikz}
\usetikzlibrary{decorations.text,calc,arrows.meta}
\usetikzlibrary{matrix,positioning,calc}
\usepackage{lipsum,adjustbox}
\tikzset{block/.style={rectangle,
                       draw,
                       fill=white,
                       rounded corners,
                       minimum height=2.5em,
                       minimum width =11em,
                       text centered, 
                       text width=11em},
         			   >=latex}

\usepackage[flushleft]{threeparttable} 

\newcolumntype{L}[1]{>{\raggedright\let\newline\\\arraybackslash\hspace{0pt}}m{#1}}
\newcolumntype{C}[1]{>{\centering\let\newline\\\arraybackslash\hspace{0pt}}m{#1}}
\newcolumntype{R}[1]{>{\raggedleft\let\newline\\\arraybackslash\hspace{0pt}}m{#1}}
\newcommand{\changed}[1]{\textcolor{black}{#1}}%

\newcommand{\cmark}{\ding{51}}%
\newcommand{\dquotes}[1]{``#1''}

\makeatletter 
\makeatletter

\definecolor{blue(pigment)}{rgb}{0.2, 0.2, 0.7}

\newif\ifworkinprogress
\workinprogresstrue

\definecolor{darkgreen}{rgb}{0, .4, 0}
\definecolor{ydc}{rgb}{0.9, .4, 0.5}

\ifworkinprogress
	\newcommand{\markus}[1]{\textcolor{darkgreen}{\textbf{[Markus] #1}}}
	\newcommand{\ms}[1]{\textcolor{darkgreen}{\textbf{[Markus] #1}}}
	\newcommand{\yashar}[1]{\textcolor{ydc}{\textbf{[Yashar] #1}}}
	\newcommand{\peter}[1]{\textcolor{red}{\textbf{[Peter] #1}}}
\else
  \newcommand{\markus}[1]{}
  \newcommand{\ms}[1]{}
  \newcommand{\yashar}[1]{}
  \newcommand{\peter}[1]{}
\fi

\definecolor{blue(pigment)}{rgb}{0.2, 0.2, 0.7}

\begin{document}

\begin{frontmatter}

\title{Content-driven Music Recommendation: Evolution, State of the Art, and Challenges}



\author[poliba]{Yashar Deldjoo\corref{mycorrespondingauthor}}
\cortext[mycorrespondingauthor]{Corresponding author}
\ead{deldjooy@acm.org}

\author[jku]{Markus Schedl}
\ead{markus.schedl@jku.at}

\author[tuw]{Peter Knees}
\ead{peter.knees@tuwien.ac.at}

\address[poliba]{Polytechnic University of Bari, SisInf Lab, Dept. of Electrical Engineering and Information Technology, Via Orabona 4, Bari, Italy}
\address[jku]{Johannes Kepler University Linz, Institute of Computational Perception (Multimedia Mining and Search Group) and Linz Institute of Technology, AI Lab (Human-centered AI Group), Linz, Austria}
\address[tuw]{TU Wien, Faculty of Informatics, Institute of Information Systems Engineering, Vienna, Austria}

\begin{abstract}
The music domain is among the most important ones for adopting recommender systems technology.
In contrast to most other recommendation domains, which predominantly rely on collaborative filtering (CF) techniques, music recommenders have traditionally embraced content-based (CB) approaches.
In the past years, music recommendation models that leverage collaborative and content data --- which we refer to as content-driven models --- have been replacing pure CF or CB models. 
In this survey, we review \changed{55} articles on content-driven music recommendation. Based on a thorough literature analysis, we first propose an onion model comprising five layers, each of which corresponds to a category of music content we identified: signal, embedded metadata, expert-generated content, user-generated content, and derivative content. We provide a detailed characterization of each category along several dimensions.
Second, we identify six overarching challenges, according to which we organize our main discussion: 
increasing recommendation diversity and novelty, 
providing transparency and explanations, 
accomplishing context-awareness, 
recommending sequences of music, 
improving scalability and efficiency, and 
alleviating cold start.
Each article addresses one or more of these challenges is categorized according to the content layers of our onion model, the article's goal(s), and main methodological choices. Furthermore, articles are discussed in temporal order to shed light on the evolution of content-driven music recommendation strategies.
Finally, we provide our personal selection of the persisting grand challenges which are still waiting to be solved in future research endeavors.
\end{abstract}

\end{frontmatter}

\section{Introduction and Motivation}\label{sec:intro}
Thanks to the recent evolution in web technologies and the availability of numerous multimedia streaming services, customers can access an incredible amount of music online for a flat rate or even free of charge. As music collections offered to the consumers of streaming services continue to expand, \changed{it becomes increasingly difficult for users to discover new content that is engaging. Even identifying a specific piece of music may equal searching for a needle in a haystack, especially when there are ambiguities or several versions of the same piece.} 

To lessen users' burden to identify interesting items (e.g., performers, releases, or single tracks), recommender systems (RSs) have emerged and evolved during the past decade.\footnote{While a performer is not an item, strictly speaking, we adopt here the common notion in RS research, according to which entities that are recommended are referred to as \textit{items}.} Such RSs aim to automatically determine and present items a user may like.
Substantially driven by commercial interests, research on RS topics has experienced a significant boost since then, in particular thanks to competitions such as the Netflix Prize\footnote{\url{http://www.netflixprize.com}} for movie recommendation~\cite{koren_etal:comp:2009} or, more recently, the ACM Recommender Systems Challenge 2018\footnote{\url{https://recsys-challenge.spotify.com} and \url{http://www.recsyschallenge.com/2018}} on automatic music playlist continuation~\cite{Schedl2018}, co-organized by Spotify.

{At present, the majority of industrial music recommender systems (MRSs) rely on usage patterns, whether implicit feedback or explicit ratings, leveraged by collaborative filtering (CF) models to compute personalized recommendations. Neighborhood models are among traditional CF classes that predict the unknown user--item preferences based on the user or item neighborhood, considering either like-minded users or similarly-rated items. More recent RSs, in contrast, tend to use model-based variants of CF, such as matrix factorization (MF), a parameterized model whose parameters are learned in the context of an optimization framework, such as Bayesian personalized ranking (BPR)~\cite{rendle2009bpr}. A commonly reported issue of MF methods is the linearity of these models. Deep neural networks (DNNs), for instance used in neural collaborative filtering (NCF)~\cite{he2017neural}, have been introduced to address this issue by modeling the non-linear relationships in data through non-linear activation functions. As a result, these models can unveil more complex relationships between users and items and can ultimately better reflect users' preferences.} 

\changed{In addition to neural techniques, recent efforts aggregate user-item historical interactions as a \textit{bipartite graph} to examine the underlying relationship between users and items. Conducting some meticulously crafted operations over the graph permits us to capture direct and/or indirect collaborative signals and inject them into the representations of users and items~\cite{deldjoo2022multimediach,wu2020graph}. According to~\cite{deldjoo2022multimediach}, modern graph-based recommendation methods are divided into three categories: random walk-based approaches~\cite{Adsorption,TripartitePropagation,Co-rank}, graph neural network-based approaches~\cite{ACF,PinSage,MMGCN, HFGN}, and graph auto-encoder approaches~\cite{GCMC,STAR-GCN,GCM}.}

{Notwithstanding their great success, a common issue with CF-based approaches is that the user profile is independent of the descriptive attributes of items users liked in the past. As a result, pure CF approaches disregard a wealth of \textit{in-domain knowledge} contained in the item content features. Considering such knowledge can yield \textit{(1)} better recommendations and \textit{(2)} enable explanations of recommendations~\cite{afchar:aimag:2021,DBLP:journals/ftir/ZhangC20,Tintarev2015}. Thus, a new generation of modern RSs try to leverage different types of content data as side information to CF models~\cite{DBLP:journals/csur/DeldjooSCP20,DBLP:journals/csur/ShiLH14}, resulting in increasing the expressive power of hybrid recommendation models, which leverage collaborative and content information. For brevity, we refer to such recommendation models as \textit{content-driven models}. Content-driven approaches to MRSs match the human way of perceiving music and are suited for explanation purposes since we most commonly like a music item because of its content. This is notwithstanding other reasons, such as affective experiences, memories, or contextual factors. Content-driven techniques also allow users to explore and extend their musical taste, leaving their comfort zone, e.g., by diversifying recommendation results. Content-drive MRSs are also valuable in cold-start situations where not enough interaction data exist to fuel CF models. In this case, an MRS could ask the user to identify their preferences concerning facets of the musical content, e.g., tempo, instrumentation, or style, and create an initial list of recommendations according to these indications. Readers are referred to~\cite{deldjoo2022multimediach,Musto2021Semantics_ch,schedl_ch} for an overview of recent advances in content-driven recommender systems.} 

Item content features can encompass a variety of information, ranging from metadata (e.g., genre, emotion, or instrumentation)~\cite{knees_schedl:tomccap:2013} to user-generated content (e.g., tags or reviews)~\cite{DBLP:journals/csur/ShiLH14} to features extracted from the audio signal directly~\cite{DBLP:journals/csur/DeldjooSCP20}; 
or semantic knowledge collected from a knowledge graph such as Wikidata\footnote{\url{https://www.wikidata.org}} or DBpedia~\cite{9216015}.\footnote{\url{https://www.dbpedia.org}} 

Building an MRS that leverages available metadata, such as the artist, album, and release year, is perhaps the most straightforward approach to creating an MRS. However, relying solely on such metadata will lead to predictable recommendations, i.e., an artist-based MRS will recommend the user songs that she would have listened to anyway, which is not particularly useful. In contrast, leveraging the expressive power of descriptive features extracted from the audio and other multimedia signals (e.g., music video clips) can generate more informed and less trivial recommendations.

{Content-based music features can be broadly categorized into \textit{(1)} high-level attributes/metadata (e.g., genre, artist, tags,  Wikidata) and \textit{(2)} low-level features, which are extracted from the core audio signal (e.g., pitch, timbre, or tempo) using audio/music signal processing and  machine learning techniques. Although the MRS community has been using these versatile types of information in various forms, tasks, and quantities, we believe there is a lack of profound understanding of the following key questions:}
\begin{itemize}
    \item {\textit{\textbf{H}ow} can we categorize, in an unambiguous manner, the various types of content features used in MRS research?}
    
    \item {\textit{\textbf{W}hich} are the main techniques to build MRS that leverage data of these different categories of content information?}
    
    \item {\textit{\textbf{W}hat} are the main challenges in MRS research and how can they be successfully approached by leveraging music content data?} 
\end{itemize}

Addressing these questions, the major contributions of this survey article are: 
\begin{itemize}
	\item We propose a hierarchical model to describe the various types of characteristics of music items along the continuum of content vs.~context descriptors (Section~\ref{sec:content_levels}). 
	\item We identify, discuss, and categorize the state-of-the-art approaches to content-driven MRS, along the dimensions of the proposed model, their goals and addressed challenges, their original contributions, their historic evolution, and their representation of content (Section~\ref{sec:challenges}).
	\item We identify and outline the grand challenges MRS research is currently still facing (Section~\ref{sec:conclusion}).
\end{itemize}

The survey is organized in the following manner: 
In Section~\ref{sec:literature}, we position this article in the context of existing surveys and overviews about related topics. We further present the methodology we adopted to identify relevant publications. 
Section~\ref{sec:content_levels} subsequently introduces our hierarchical model to describe the different levels of content, along the continuum between purely audio-based descriptors (\dquotes{content} in a narrow interpretation) and external data that is not part of the original signal, nevertheless descriptive or related to a music item. \changed{Section~\ref{sec:methods_MRS} provides an overview of the most prominent recommendation models and strategies to exploit one or more layer(s) of the onion model as rich additional information beyond CF signals.} Section~\ref{sec:challenges} represents the core part of this survey. 
We categorize the reviewed articles according to their \textit{content category} (as defined in our model presented in Section~\ref{sec:content_levels}), their \textit{content representation}, and the specific \textit{audio features} in cases where the audio signal is used. On the highest level, we categorize reviewed literature according to the main challenges and goals addressed, and we structure this section accordingly.
Section~\ref{sec:conclusion} concludes the article and elaborates on the open grand challenges we believe MRS research is facing. 

Table~\ref{tbl:abbr} provides a list of abbreviations used throughout the article.

\begin{table}[t]
\centering
\caption{\label{tbl:abbr}List of abbreviations used throughout the survey.} 
\scalebox{0.80}{
 \begin{tabular}{l|l}
\toprule
\multicolumn{1}{c}{\textbf{Abbreviation}} & \multicolumn{1}{c}{\textbf{Term}} \\
\bottomrule
ALS & Alternating least squares \\ \hline
APC & Automatic playlist continuation \\ \hline
APG & Automatic playlist generation \\ \hline
CA & Context-aware \\ \hline
CBF & Content-based filtering \\ \hline
CF & Collaborative filtering \\ \hline
CS & Cold start \\ \hline
CD-RS & Cross-domain recommender system\\ \hline
\changed{CNN} & \changed{Convolutional neural network} \\ \hline
CV & Computer vision \\ \hline
DL & Deep learning \\ \hline
DNN & Deep neural network \\ \hline
DWCH & Daubechies wavelet coefficient histograms \\  \hline
E2E & End-to-end \\ \hline
FM & Factorization machine \\ \hline
\changed{GCN} & \changed{Graph convolutional network} \\ \hline
\changed{GNN} & \changed{Graph neural network} \\ \hline
GMM & Gaussian mixture model \\ \hline
IR & Information retrieval \\ \hline
KNN & K-nearest neighbors \\ \hline
LFM & Latent factor model \\ \hline
LSTM & Long short-term memory \\ \hline
MAP & Mean average precision \\ \hline
MER & Music emotion recognition \\ \hline
MIR & Music information retrieval \\ \hline
MF & Matrix factorization \\ \hline
MFCC & Mel frequency cepstral coefficents \\ \hline
ML & Machine learning \\ \hline
MRS & Music recommender system \\ \hline
MLR & Metric learning to rank \\ \hline
NDCG & Nomalized discounted cumulative gain \\ \hline
RMSE & Root-mean-square error \\ \hline
\changed{RNN} & \changed{Recurrent neural network} \\ \hline
RS & Recommender system \\ \hline
SAR & Sequence-aware recommender \\ \hline
\changed{SNN} & \changed{Siamese neural network} \\ \hline
SM & Social media \\ \hline
SN & Social network \\ \hline
SVR & Support vector regression \\ \hline
TF-IDF & term frequency-inverse document frequency \\ \hline
URM & User rating matrix \\ \hline
VAE & Variational autoencoder \\ \hline
ZCR & Zero crossing rate \\
\bottomrule
\end{tabular}
} 
\end{table}

\section{Survey Context and Literature Search Methodology}\label{sec:literature}

\subsection{Related Surveys}
To put this survey into context, we identified the following detailed overview and survey articles and explain in which ways the article at hand connects to and extends earlier work.
In an earlier article~\cite{knees_schedl:tomccap:2013}, we presented a survey of music similarity and recommendation from music context data, therefore being agnostic of audio information. 
We discussed works that adopt the strategy of turning the music similarity task into a text similarity task by considering sources such as music-related web pages, microblogs, and tags.
Dating back to 2013, that survey does not account for recent advances made possible, for instance, through deep learning, sophisticated listener modeling including alternative concepts such as personality or culture, or user-centric evaluation schemes.
The survey at hand is also complementary to~\cite{knees_schedl:tomccap:2013} in that it focuses on content descriptors instead of contextual data, targets both the multimedia and recommender systems communities, and organizes discussed research according to several identified recent challenges, which are addressed by the selected papers.

Kaminskas et al.~\cite{kaminskas2012contextual} provide a survey on context-aware music recommendation and retrieval. In addition to the traditionally exploited situational signals, such as time and location, the authors discuss the use of affective cues, such as mood and emotion, and how they can be integrated into music recommendation applications.
While the topic of music recommendation is the same as in the survey at hand, Kaminskas et al.~\cite{kaminskas2012contextual} solely review approaches that consider contextual information. In contrast, our work focuses on research that exploits content data and information derived from there.

Quadrana et al.~\cite{quadrana2018sequence} give an overview of sequence-aware recommender systems. While this is a very recent survey, the authors do not focus on the music domain in particular. However, producing a meaningful sequence of music tracks is a critical task in MRS research and is often referred to as automatic (music) playlist generation (APG). The authors, therefore, also review several approaches for APG that only employ methods of sequence learning (e.g., frequent pattern mining, recurrent neural networks, and Markov chains). Content-based methods are discussed to a lesser extent. 

In an earlier survey, Bonnin et al.~\cite{Bonnin:2014:AGM:2658850.2652481} particularly discuss APG. While focusing on research methods that generate sequences of songs, that survey also discusses basics in content-based similarity extracted from audio or textual features and their use for APG. 
While we also identify APG as one of the main challenges in content-based MRS, we extensively discuss in the survey at hand how content features can be incorporated into the sequential recommendation process, and we also provide an up-to-date review.

Acknowledging the recent trend of using deep learning techniques also in the field of recommender systems, Zhang et al.~\cite{zhang2017deep} review respective approaches in various domains, including image, point of interest, news, and hashtag recommendation.
The authors also mention works that exploit convolutional neural networks trained from audio spectrograms and deep collaborative filtering strategies for music recommendation. However, music recommendation plays only a minor role in that survey.

{Murthy et al.~\cite{Murthy:2018:CMI:3212709.3177849} present a survey on content-based music information retrieval (CB-MIR). The authors review the common 
acoustic features used in MIR (harmony, pitch, rhythm) and major MIR-related tasks such as artist identification, genre classification, or 
emotion recognition. 
While the discussion of content-based music features is relevant to the survey at hand, 
the authors do not focus on MRS tasks and challenges. 
}

{Finally, Deldjoo et al.~\cite{DBLP:journals/csur/DeldjooSCP20} present a survey on RS leveraging multimedia content, i.e., audio, visual, or textual content, and combinations thereof. That survey provides an overview of various RS techniques based on classical model-based CF leveraging latent content factor and recent models using deep neural networks and graph learning.
A small part of that survey is also dedicated to MRSs that exploit multimedia content.  
However, only a few works are discussed, and a comprehensive study of the types, roles, and significance of content in MRS, as offered in the survey at hand, is missing. The authors provide a more in-depth survey of algorithms and strategies capable of utilizing multimedia content in \cite{deldjoo2022multimediach}, which may also be employed for MRS; hence, this study is pertinent to the current survey.}

\subsection{Selection of Relevant Publications}\label{sec:literature_search}

\begin{figure}
\begin{subfigure}{.45\textwidth}
  \centering  
\includegraphics[width=0.930\linewidth]{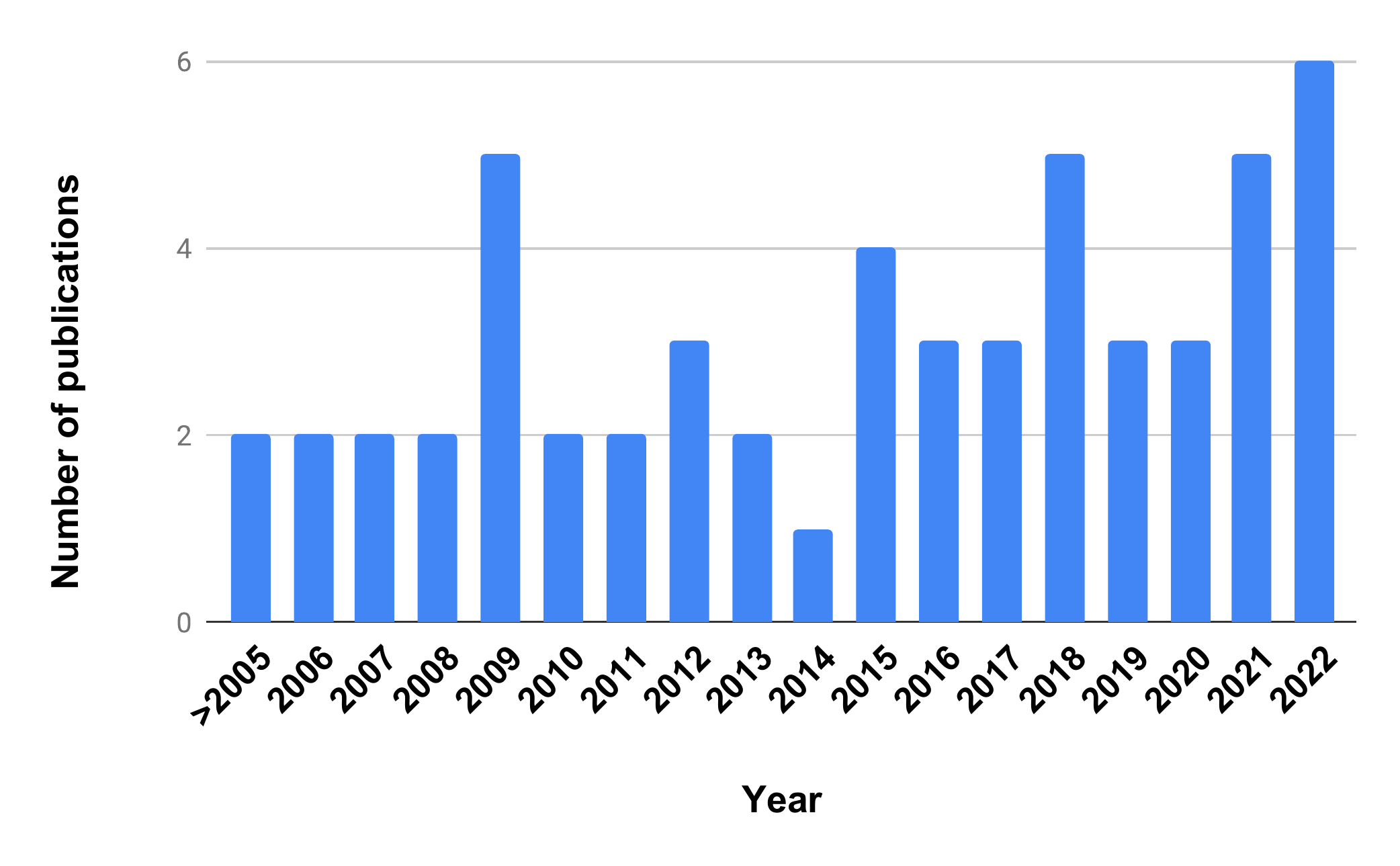}
  \caption{Number of publications over the years}
  \label{fig:sub-first1}
\end{subfigure}
\begin{subfigure}{.45\textwidth}
  \centering
  \includegraphics[width=.99\linewidth]{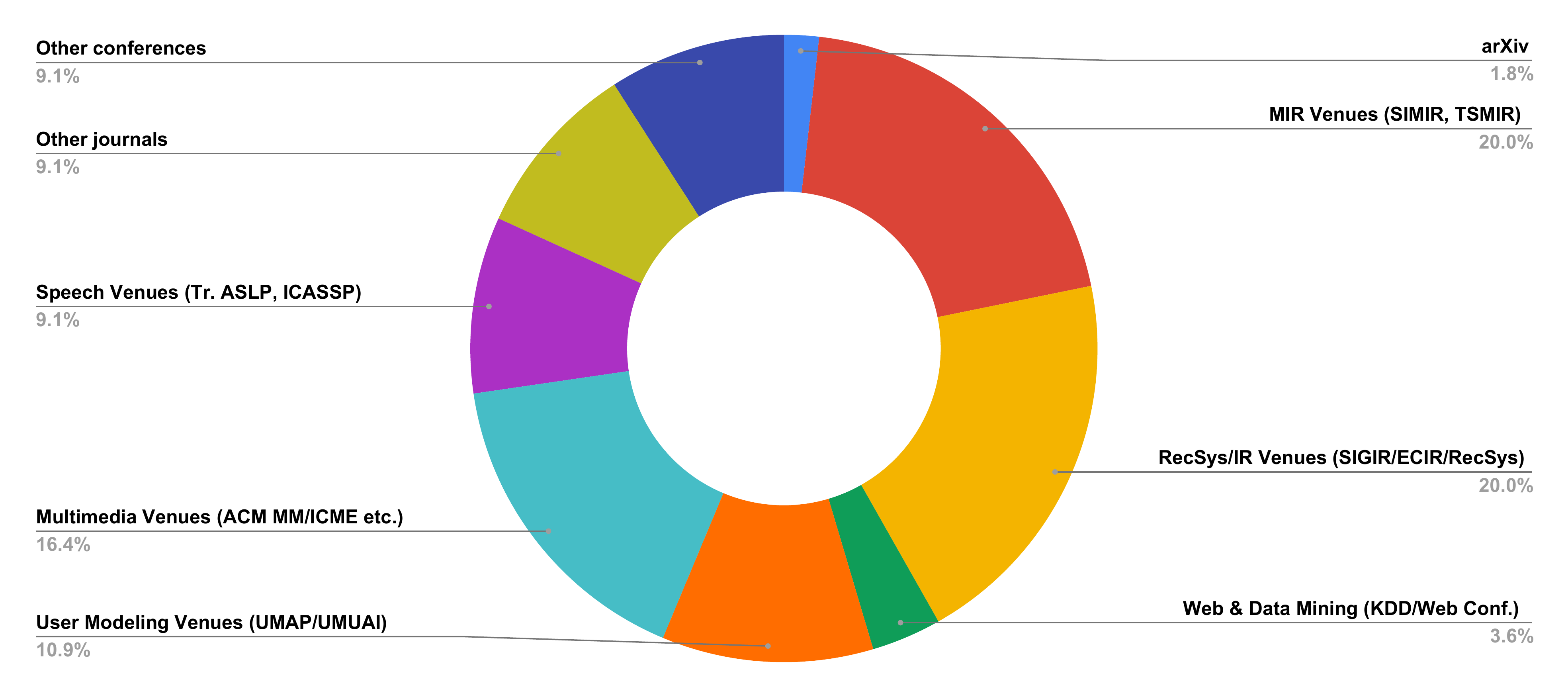}  
  \caption{Distribution based on the venue}
  \label{fig:sub-second1}
\end{subfigure}
\caption{Statistics of publications related to content-driven MRSs with respect to publication year and venue.}
\label{fig:stats}
\end{figure}

To identify the relevant publications that form state of the art, we applied a two-stage search strategy. First, we considered relevant research works from top conferences in the fields of recommender systems, multimedia processing and retrieval, {and music/audio signal analysis}, including the ACM Conference on Recommender Systems (RecSys),\footnote{\url{http://www.recsys.acm.org}} ACM Conference on Multimedia,\footnote{\url{http://www.acmmm.org}} the International ACM SIGIR Conference on Research and Development in Information Retrieval (SIGIR),\changed{\footnote{\url{http://www.sigir.org}} 
ACM Conference on User Modeling, Adaptation and Personalization (UMAP),\footnote{\url{https://www.um.org}}, and the Journal of User Modeling and User-Adapted Interaction (UMUAI)}, 
and the International Society for Music Information Retrieval Conference (ISMIR).\footnote{\url{http://www.ismir.net}} 
We filtered for papers using search terms such as \dquotes{audio content}, \dquotes{acoustic content}, \dquotes{music recommender system}, and variants of those.
We collected all resulting papers since the first edition of the RecSys conference and ISMIR, and those published between 2010 and 2020 for the other venues.\footnote{Since the conferences other than the ACM Conference on Recommender Systems have a much longer history, we refrained from searching back until the first edition of these venues, from avoiding including possibly outdated works.}

Bearing in mind that many other venues on (multimedia) information retrieval and multimedia signal processing also publish related works, we also gathered  a number of related publications by searching directly through Google Scholar\footnote{\url{https://scholar.google.com}} and Dblp,\footnote{\url{https://dblp.org}} and filtering for the name of the most important journals, 
including the IEEE Transactions on Multimedia (TR-MM),\footnote{\url{http://ieeexplore.ieee.org/xpl/RecentIssue.jsp?punumber=6046}} Multimedia Tools and Applications (MTAP),\footnote{\url{http://www.springer.com/computer/information+systems+and+applications/journal/11042}} ACM Transactions on Intelligent Systems and Technology (TIST),\footnote{\url{http://www.tist.acm.org}} and IEEE Transactions on Audio, Speech, and Language Processing (TR-ASLP),\footnote{\url{http://www.ieeexplore.ieee.org/xpl/RecentIssue.jsp?punumber=10376}} and Transactions of the International Society of Music Information Retrieval (TISMIR).\footnote{\url{https://transactions.ismir.net}}
Finally, we searched Google Scholar (using the search terms mentioned above) to obtain a number of papers from other venues. 
We mostly turned our attention to conference and journal publications and to a lesser extent to workshop publications.

After having read the publications identified as described above, in a second step, we looked for the closest works to each publication, by considering the citations in the target publication as well as the related work option in Google Scholar. The venues for these were not limited to the ones mentioned above. Applying this multi-stage approach yielded a total of \changed{!!!47!!!} publications. {Figure~\ref{fig:stats} shows the distribution of reviewed papers based on publication year and venue. It can be noticed in Figure~\ref{fig:sub-first1} that the 
number of papers shows a slight upward trend starting in the mid 2010s, where the last years coincide with the remarkable rise of deep learning (DL).

In addition, we can see in Figure~\ref{fig:sub-second1} that most of the papers are published in top conferences and journals, which we categorize into \changed{MIR-oriented venues  (e.g., ISMIR and its journal TISMIR), multimedia venues (i.e., ACM MM/ICME/MMSys/IJMIR),   information retrieval and recommender systems conferences (e.g., SIGIR and RecSys), and Web and Data Mining Venues (e.g., KDD, Web) among others.} We do not claim that the list of publications identified by our selection approach is exhaustive. Nevertheless, we strongly believe this survey provides an overall picture of the recent advances, trends, and challenges in the field, and useful insights for researchers and practitioners in academia and industry.



\section{Levels of Content}\label{sec:content_levels}
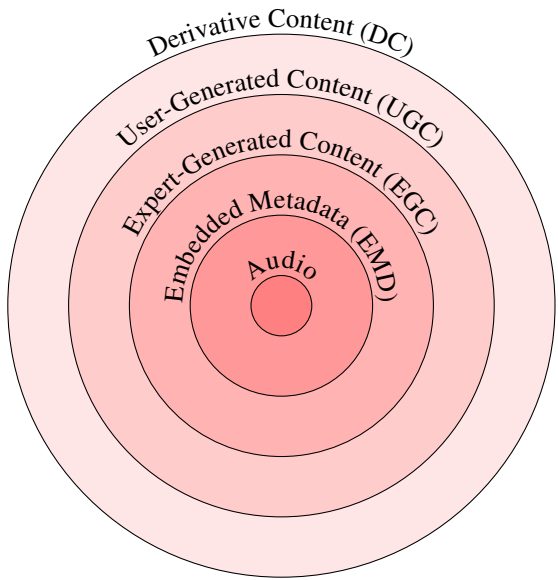
\begin{figure}[!th]
\centering
\begin{tikzpicture}
\coordinate (O) at (0,0);

\begin{scope}[xshift=6cm]
\coordinate (O) at (0,0);
\draw[fill=red!10] (O) circle (3.6);
\draw[fill=red!20] (O) circle (2.8);
\draw[fill=red!30] (O) circle (2);
\draw[fill=red!40] (O) circle (1.2);
\draw[fill=red!50] (O) circle (0.4);

\draw[decoration={text along path,reverse path,text align={align=center},text={Audio}},decorate] (0.5,0) arc (0:180:0.5);
\draw[decoration={text along path,reverse path,text align={align=center},text={Embedded Metadata (EMD)}},decorate] (1.3,0) arc (0:180:1.3);
\draw[decoration={text along path,reverse path,text align={align=center},text={Expert-Generated Content (EGC)}},decorate] (2.1,0) arc (0:180:2.1);
\draw[decoration={text along path,reverse path,text align={align=center},text={User-Generated Content (UGC)}},decorate] (2.9,0) arc (0:180:2.9);
\draw[decoration={text along path,reverse path,text align={align=center},text={Derivative Content (DC)}},decorate] (3.7,0) arc (0:180:3.7);
\end{scope}
\end{tikzpicture}
\caption{\label{fig:onion}
An ``onion model'' of content in the music domain. 
Layers of content are added to the audio signal progressively over time, reflected in the order from inner to outer layers. 
This also reflects a continuum from strictly objective and numeric item descriptors (inner) to more subjective and semantically charged content data stemming from cultural practices of dealing with music (outer).
In terms of content properties, similarly from inner to outer layers, we see a shift from data sources not suffering from cold start problems and providing credible and trustworthy information in the available data, to data sources more prone to cold start, noise, and errors, therefore requiring larger volumes of data to become a more credible source.} 
\end{figure}

Within the recommender systems research community, the term \emph{content} can be found to refer to a variety of data sources that can be accessed to derive features describing the items to be recommended. 
Examples of such sources are song attributes and user tags for music, e.g.,~\cite{Hariri:2012:CMR:2365952.2365979} or the synopsis of a movie, e.g.,~\cite{Melville:2002:CCF:777092.777124}. 
In contrast to this, traditionally, retrieval-oriented research communities, such as MIR, refer to the content of an item only if features are directly extracted from the digital representation of the item, e.g., the audio signal, and refer to related sources as metadata or context data, cf.~\cite{knees_schedl:tomccap:2013}.
In this work, we adopt the recommender systems community's broader understanding.
Hence, for music, content refers not only to an audio signal but also to various types of metadata that describe a musical item on different levels and from different perspectives.
This ranges from bibliographical records and identifiers connected to individual recordings to semantic expert annotations and user-generated and derivative data. 

In order to structure available sources of content, we devise a new hierarchical model illustrated in Figure~\ref{fig:onion}. This ``onion model'' consists of several layers of content categories, starting with the audio signal at its core and gradually adding layers of content that exhibit higher semantic valence and higher subjectivity as this data relates to musical practices and music usage as cultural artifacts.
The identified layers are described in the following from most-inner to most-outer: +

\begin{itemize}
\item \textbf{Audio}: This often lossy-encoded digital representation of content aims to reconstruct a recorded (or digitally produced) acoustic signal.
The encoded information is, therefore, only the physical signal without any semantics.  
The semantic dimension is created within the listener when perceiving the audio. Automatic audio analysis algorithms aim to model some of these processes and represent perceived dimensions by extracting content features, which can be used in the recommendation process~\cite{schedl_etal:fntir:2014}. Typical examples of content descriptors on this level are time-domain descriptors, e.g., loudness or zero-crossing-rate; spectral descriptors, e.g., rolloff or MFCCs; tonal descriptors, e.g., melody or chroma; and rhythm descriptors, e.g., onsets or beats; cf.~\cite{Bogdanov:2013:EOL:2502081.2502229} for an overview.  
\item \textbf{Embedded Metadata (EMD)}:
This layer of content is very closely connected to the music item. It refers to descriptive and technical metadata such as record identifiers and accompanying data, comprising information on, e.g., artist name, track title, album title, lyrics, label, year of recording, producer, or production settings, and even other multimedia data such as album cover artwork. We furthermore consider the ``official'' video clip of a music piece to belong to this layer of content.
These bibliographical records are often directly embedded in the digital container of the audio (e.g., ID3 tags in MP3 files or RIFF and BWF information in Wave files) 
or externally recorded in separate files or web-accessible databases (e.g., MusicBrainz, Discogs, Music Ontology~\cite{raimond_etal:ismir:2007}), cf. Corthaut et al.~\cite{nik_corthaut_2008_1415244}, and linked to the audio using unique identifiers such as audio fingerprints, e.g., used by Shazam~\cite{wang:2003:ismir}.\footnote{URLs for reference: \url{https://musicbrainz.org}, \url{https://www.discogs.com}, \url{http://musicontology.com}}
Besides providing elementary content descriptors useful in recommendation, e.g., genre, embedded metadata often works as a bridge to the higher layers, e.g., to link audio content to the expert- and user-generated and derivative content.

\item \textbf{Expert-Generated Content (EGC)}: 
A richer and more semantic, albeit the more opinionated type of content, is music metadata assigned from a trained listener's or musicological perspective.
This comprises attributes such as genre, style, or mood; contextualizes music in terms of era, origin, and trends (often defined or adapted only in retrospect), and provides more detailed descriptors of the content (instrumentation, sound qualities, vocal properties, structural properties, lyrical content, song message, etc.) or performing artists (e.g., biographies).
Examples of sources of this type of content are the Music Genome Project~\cite{prockup_etal:ismir:2015}, Allmusic, and Wikipedia.\footnote{URLs: \url{https://www.allmusic.com}, \url{https://www.wikipedia.org}}
This type of content data is typically of very high quality, however, approaches to obtain this data are less scalable and might exhibit biases such as expert preferences and cultural viewpoints and priorities.
\changed{Please note that textual descriptors at this content layer (together with those at the UGC layer) is often used as target in the task of music auto-tagging~\cite{DBLP:conf/icassp/IbrahimREPR20,DBLP:journals/tmm/LinC21,DBLP:conf/eusipco/FerraroBSJY20}. Auto-tagging leverages audio content to predict a set of labels for a given song; it can therefore be considered a multi-class multi-label classification task.}

\item \textbf{User-Generated Content (UGC)}:
Moens et al.~\cite{moens2014mining} define user-generated content as content ``published on a publicly accessible Web site or a page on a social networking site that is accessible to a select group of people'' comprising ``blogs, wikis, discussion forums, posts, chats, tweets, podcasting, pins, digital images, video, audio files, and other forms of media that users created''.
Typical examples of UCG in music information retrieval --- referred to as ``community metadata'' by Whitman et al.~\cite{whitman_lawrence:icmc:2002} --- are tags (created and shared, for instance, on Last.fm, e.g.~\cite{lamere:nmr:2008,levy_sandler:jnmr:2008}), reviews (e.g., on Amazon or Pitchfork, cf.~\cite{downie_hu:jcdl:2006}), song explanations (e.g. on Genius~\cite{fields_rhodes:ismir:2016}), playlists (e.g.~\cite{mcfee_lanckriet:ismir:2012}), tagged tweets (e.g.~\cite{hauger_schedl:cbmi:2016}), or general web pages (e.g.~\cite{knees_etal:ismir:2004}).\footnote{URLs: \url{https://www.last.fm}, \url{https://www.amazon.com}, \url{https://pitchfork.com}, \url{https://genius.com}}
As pointed out by Moens et al.~\cite{moens2014mining} --- and also relevant for music data --- UGC ``comes in a variety of languages and modalities such as text, image, video, and location-based information, and the corresponding metadata''.
However, it needs to be stated that here, UGC is itself not the content that is being recommended (as is the case, e.g., for videos, photos, posts, etc.) but content that is analyzed in order to recommend the music it is associated with, i.e., the core of this onion model.

\item \textbf{Derivative Content (DC)}: 
Derivative content refers to new works based on the original content from the inner layers of the model, still providing information relevant for content analysis and recommendation.
This comprises \emph{re-use} of original content such as in remixes, mash-ups, cover versions, or parodies and \emph{re-purposing} (re-contextualization) of content, e.g., to score movies, videos, or ads, yielding a different perception of the original content.
Derivative content is often performed and/or produced by amateurs, cf.~UGC, and prominently featured on and obtainable from platforms such as YouTube, ccMixter (cf.~\cite{cheliotis_yew:ct:2009}), or the Japanese Niconico (cf.~\cite{hamasaki_etal:uxtv:2008}).\footnote{URLs: \url{https://www.youtube.com}, \url{http://ccmixter.org}, \url{http://www.nicovideo.jp}}
While this content layer has, so far, not received as much attention for content-based music recommendation as to the underlying layers, we believe that this will be an important direction in future work on music recommendation, cf.~\cite{tsukuda_etal:cikm:2016}.

\end{itemize}

When considering this onion model, several partly interconnected trends and continua across layers can be observed:
\begin{description}
\item [Temporal component:] 
Starting from the core, i.e., the audio, further layers are added progressively over time. 
From the sound production, data is enriched with embedded metadata before being released, first to experts (e.g., journalists; EGC), later to the general public (leading to UGC).
Derivative content is contributed the latest.\footnote{Note that this is not a strict process as outer layers can inform and impact inner layers.}

\item [Semantics:] 
While audio data --- and the features extracted from it --- comprise \emph{numeric values} to represent the signal, outer layers increasingly contain \emph{semantically charged} data. 
Rather than a representation of music, this refers to artifacts stemming from cultural practices of dealing with music.

\item [Subjectivity:] 
At the same time, we can also observe a shift from an objective representation (the signal) to data consisting of subjective categories and impressions.
While this subjectivity and individuality in music perception are central to the semantic dimension, this makes it more challenging to derive general feature representations that refer to an item without introducing the dimension of the user.
Conversely, subjectively biased data at a large scale might provide a better basis for making recommendations than objective features, cf.~\cite{slaney:ieeemm:2011}. 

\item [Cultural context:] 
Similarly, while the core of the model contains pure content data, layers towards the outside exhibit more contextual and culturally-driven information,  providing more realistic and typical data, 
more appropriate for culture-specific recommendation, cf.~\cite{zangerle_etal:umap:2018}.

\item [Community:] 
The presence of outer layers depends on the presence of users creating and contributing related content.
Production of the audio and the embedded metadata is usually carried out by a small group of people, creation of UGC and DC typically requires a larger community responding to the music.

\item [Cold start:] 
As a consequence, a central challenge of recommender systems, namely the cold start problem for item information, increases towards the outer layers due to increasing data scarcity within individual sources.
While the automatic content analysis does not suffer from a cold start at all, the remaining layers might not be present (or inaccessible due to missing or mismatching embedded metadata).
Cold start at the more semantic layers shows to be a significant motivation for exploiting audio analysis, e.g., in hybrid recommender systems (see Section~\ref{sec:challenges}).

\item [Data diversity:] 
While the core of the model represents the item through one audio file, the remaining layers can consist of a broader spectrum of data from multiple sources.
This spectrum increases the volume of data that needs to be accessed and processed and the diversity of information, incorporating different aspects and viewpoints. 

\item [Quality:] 
With diverse data from several sources, we can also identify an increase of noisiness in that data.
With the core representing --- literally --- the pure signal, the inclusion of external information leads to the introduction of erroneous and inconsistent data.

\item [Credibility:] 
The increase of noise also leads to less credibility of the data sources at the outer layers, as these are also susceptible to manipulation.
While extracting features directly from the audio signal provides trust in the information extracted from it, outer layer features strongly depend on external sources, which are frequently prone to hacking and vandalism. An infamous example is Paris Hilton being tagged as ``brutal death metal'' on Last.fm, cf.~\cite{lamere:nmr:2008}. Therefore, a larger amount of data, ideally from independent sources, is needed to achieve a similar level of trust and credibility. 

\end{description}

Different layers of content exhibit different properties, advantages, and disadvantages; they have been used to address different challenges of MRSs, as described next. Recent research by Moscati et al. \cite{moscati2022music4all}  publishes a dataset named Music4All-onion that attempts to systematically evaluate the influence of various degrees of content in content-driven MRS.

\changed{\section{Music Recommendation Exploiting Onion-Model Content as Side Information}\label{sec:methods_MRS}}

\changed{We can generally categorize approaches that exploit item content based on one or more levels of the onion model, as described below:}

\changed{
\begin{itemize}
    \item Classical model-based CF
    \item Deep neural models
    \item Graph-based models
    \item Other strategies
\end{itemize}}

\changed{For describing each of these strategies to integrate rich item content as side information, we first provide a brief description for each class of techniques, followed by MRS-specific examples from the literature.}
\vspace{-2mm}
\changed{\subsection{Classical model-based CF}}

\changed{The simplest way to integrate audio content into standard collaborative RSs is to use a classical approach, which refers to 
models that are not based on more modern neural or graph architectures. In essence, CBF models underperform CF models~\cite{DBLP:journals/csur/DeldjooSCP20}, particularly when content is represented by the onion model's innermost layer (i.e., Audio). As tags convey both user folksonomy signals and content-based information, outer layers such as UGC can fuel pure CBF models, particularly in cold-start scenarios where CF signals are rare. A good example of such work in the music domain exploiting pure CBF is~\cite{oramas2017deep}, examine the effect of several types of content on cold-start music recommendation, focusing on tags in pure CBF using the item-KNN technique.}

\changed{According to~\cite{deldjoo2022multimediach}, \textit{extended memory-based CF} and  \textit{extended model-based CF} are two paradigms for integrating content as side information, and this may be applied to any sort of content described by the onion model. For the first category, item-KNN might be modified to incorporate similarity based on interaction and content data. Several model-based strategies could be utilized for the latter category, such as those extending MF models~\cite{DBLP:journals/umuai/DeldjooDCECSIC19}, based on the factorization machines (FM), or those using contextualized sparse linear methods (cSILM).}

\changed{In~\cite{DBLP:journals/umuai/DeldjooDCECSIC19}, the authors describe a two-stage recommendation strategy for cold-item recommendations by learning the weights of a feature weighting CBF algorithm using interaction data and content-information (in two phases separately). The authors apply their methods to a variety of multimedia content, including audio content, and evaluate the quality of cold-item recommendations. Although this strategy has been used to evaluate a movie recommender system, it can also be utilized in music recommendation engines.}

\changed{Ning and Karypis~\cite{10.1145/2365952.2365983} offer different extensions to the well-known SLIM model, which is an effective CF model that learns sparse aggregation coefficients, i.e., the users' scores are modeled as a sparse aggregate of their item interactions. Contextual-SLIM is an extension that shares the characteristic of being a linear model developed from collaborative interaction data and constrained by context (e.g., item content) as side information. In the music domain,~\cite{zheng2015similarity,sassi2021morec} apply this technique to build a MRS.} 

\changed{It should be noted that typically, these approaches have employed textual side data such as reviews, tags, and metadata, which represent a single, relatively simple-to-process modality. On the other hand, audio content often needs highly effective and scalable algorithms. For example, in Factorization Machines (FMs) while these techniques work as intended for low-dimensional feature vectors, they may present scaling complications if high-dimensional feature vectors such as those obtained based on audio content are considered, mainly if higher-order interactions are accounted for in FMS.} 

\changed{\subsection{Deep neural models}}

\changed{Deep learning technologies can greatly improve MRSs, both in terms of feature extraction and the actual recommendation model. The modular design of deep learning algorithms enables the simple integration of diverse data sources, such as the content characteristics, the event history (previous behavior) of the users, and (bandit) feedback received from interactions with the recommender~\cite{deldjoo2022multimediach}.} 

\changed{In the great majority of neural models used to simulate the interaction between user and audio (but in general multimedia) content, the representations learnt for audio content by neural techniques serve as latent feature vectors. To compute recommendations, for instance, one can compute the dot product of the content representation and a learnt user feature vector, utilize the user and content features in a classifier to estimate a preference, or utilize an attention mechanism for personalized content. To combine audio and interaction representations, different fusion operations, aggregation, and concatenation to mapping become possible if the models employ both CF-based and content-based representations.}

\changed{Learned representations are important to the success of any model-based method, including neural approaches, that will ultimately drive the most useful recommendations. \textit{Multi-phase} and \textit{end-to-end} learning are both viable options for representing data in a neural architecture. Multi-phase learning requires fewer resources and is more easily applicable at scale, while end-to-end learning learns task-specific representations and is, thus, more accurate.}

\changed{The work by van den Oord et al.~\cite{NIPS2013_5004} is one of the earliest and most frequently cited deep learning-based approach to CBF in the music domain. The authors train a convolutional neural network (CNN) to predict user–item latent variables from CF matrix factorization data. This CNN can then infer these latent representations from audio alone. Recent approaches for item content representations based on deep learning include~\cite{oramas2017deep,lin2018heterogeneous,DBLP:conf/recsys/SachdevaGP18,10.1145/3320435.3320445,vall2019feature}}.

\changed{\subsection{Graph-based approaches}
Recent research has demonstrated that graph-neural networks (GNNs) are highly effective in analyzing graph-structured data and they are now commonly used for recommendation tasks. As a result, there has been a surge of GNN-based approaches to learning representations that encode not only structural information about the graph but also the content information of graph nodes. In the following, we provide an overview of recent developments in representation learning on GNNs with content data. It should be noted that while~\cite{weng2022graph,wang2017music} explore GNNs for the music recommendation task, ~\cite{wei2020graph,cai2021heterogeneous,tao2020mgat} extend the task for the recommendations for multimedia content where the audio is regarded as content. Despite being designed to produce recommendations of distinct sorts of information items (video, music), we find that such multimedia recommender systems are relevant to the current survey since they both use audio content in the graph structure to generate recommendations.}

\changed{In contrast to conventional GNNs, which rely purely on collaborative signals for representation learning, certain approaches incorporate content information to increase the GNN's capacity to construct embeddings of nodes that mix graph structure with node content. Graph-based Attentive Sequential model with Metadata (GASM) proposed by Weng et al.~\cite{weng2022graph} initially generates a directed listening graph based on metadata and then learns the representations of various types of nodes (user, item, singer, album) by GNN for prediction. Heterogeneous Information Graph Embedding method (HIGE) proposed by Wang et al.~\cite{wang2017music} improves graph representation learning by using more information to meet the performance limitation of conventional music selection. It constructs a heterogeneous information tree to store both user-music interactions and music-playing sequences. Multi-modal Graph Convolution Network (MMGCN) for
Personalized Recommendation of Micro-video proposed by Wei et al.~\cite{MMGCN} extends content across many modalities, including audio, by constructing graphs and conducting graph convolutional operations to capture modality-specific user preferences and simultaneously distilling item representations. Thus, the learned user representation can thoroughly reflect users' item-related interests. Following MMGCN, Wei et al.~\cite{wei2020graph} propose GRCN, focusing on adapting the topology of the interaction graph to identify and eliminate probable false-positive edges, hence resolving the noisy content in representation learning. Through a modality-aware heterogeneous information graph, or Heterogeneous
Hierarchical Feature Aggregation Network (HHFAN), Cai et al.~\cite{cai2021heterogeneous} investigate the highly complex relationships between users, items, and related content information of multiple modalities (e.g., visual, acoustic, and textual), and consequently learns higher-quality user and item embeddings for a personalized recommendation. Using the gated attention mechanism, Multimodal Graph Attention Network for Recommendation (MGAT) proposed by Tao et al.~\cite{tao2020mgat} propagates information within multimodal interaction networks while adaptively assigning priority to distinct modalities. Therefore, it can separate the user's preference by modality and deliver a more precise recommendation.}

\subsection{Other approaches}
\changed{Finally, we should highlight that for MRSs involving audio content, given the enormous number of items often found in music streaming services, especially in computational- or storage-limited applications, it may be possible to perform the recommendation in \textit{two phases}. The first phase employs a CF model that generates a list of candidate recommendations based on collaborative data, narrowing the search space from millions to hundreds of objects. Using audio content side data, a CBF component reorders these candidates and ultimately selects tens of items to recommend to the target users. In mobile applications, for instance, the initial stage may be pre-fetched on user devices depending on previous interactions~\cite{koch2017vfetch,Koch:2017:PCM:3083187.3083197}.}

\section{Main Challenges}\label{sec:challenges}

In the following, we provide a discussion of relevant research, as identified according to the methodology detailed in Section~\ref{sec:literature_search}. We carefully reviewed all articles and categorized them according to their main challenges and goals addressed. In total, we identified six challenges and, for each of them, several goals. To provide a quick overview of the presented works, each of the following sections includes a table showing the goals approached and major methodology adopted, as well as the level(s) of content leveraged, according to the proposed onion model. \changed{By categorizing all evaluated papers according to the layer(s) of the onion model they employ, we determine that 68.3\% of the onion level content is audio, 32.7\% EMD, 40.0\% EGC, 52.7\% UGC, and 1.8\% DC.\footnote{\changed{Note that the aggregate of percentages does not equal 100, as the majority of papers use multiple content-level features.}} While EGC, UGC, and Audio have the largest proportion of applications, the use of more audio-level characteristics has increased in recent years due to the development of complex DNNs and graph-learning methods.}

\begin{enumerate}
     \item \textbf{Increasing recommendation diversity and novelty (Table~\ref{tab:overview_diversity}):} 
     {Works belonging to this category aim at \textit{improving the user experience} by making the recommendation list more diverse and/or including more items unknown to the user. While achieving this goal often comes at the expense of lower accuracy, it helps users discover relevant, at the same time unexpected, music items, therefore ideally fostering serendipitous encounters with music.
     Since low diversity in content-based MRS is often the result of the ``hubness'' problem, i.e., some items appear in the top-N recommendation list of many users without being similar to the users' profiles, we also discuss corresponding work in this category.
     An overview of all approaches discussed in this category is provided in Table~\ref{tab:overview_diversity}.}
     
     \item \textbf{Providing transparency and explanations (Table~\ref{tab:overview_transparency}):} 
     To provide the user with insights about the music recommendations made, various works resort to content information on different levels. We, therefore, review research that creates and visualizes different explanations based on the users' or items' \textit{neighbors}, on \textit{content features}, on \textit{context}, and \textit{audio}. 
      
     \item \textbf{Accomplishing context-awareness (Table~\ref{tab:overview_context-awareness}):} Integrating contextual information into MRSs can yield more accurate recommendations, thereby enhancing the user experience.
     However, which type of context to consider, how to acquire it, and how to properly integrate it into the recommendation algorithm 
     need careful investigation. Here, we organize our review according to the major categories of context factors we identified in the relevant literature: \textit{spatial context}, \textit{affective context}, and \textit{social context}.

     \item \textbf{Recommending sequences of music (Table~\ref{tab:overview_sequences}):} {Research belonging to this category addresses the task of recommending a meaningful sequence of music items, e.g., for automatically generating playlists or extending existing ones. A key challenge here is to ensure coherent transitions between consecutive tracks. Content-driven approaches incorporate semantic descriptors, therefore playing a crucial role in corresponding research.} 

     \item \textbf{Improving scalability and efficiency (Table~\ref{tab:overview_scalability}):}
     The aim of this category of articles is to make algorithms that scale to large amounts of data, e.g., content features or number of users. They typically approach this goal by increasing the algorithms' \textit{computational efficiency} or by \textit{caching} content such as audio or video (in case of music videos).
     
     \item \textbf{Alleviating cold start (Table~\ref{tab:overview_coldstart}):} {A growing number of music tracks are released every day. 
     For MRSs, this poses the challenge of how to recommend new artists or songs when user-behavioral data are scarce. Research belonging to this category, therefore, includes 
     works that leverage content features to address the main cold start challenges, i.e., \textit{data sparsity}, \textit{new item}, and \textit{new user} problems.}

\end{enumerate}
 
In the subsequent sections, we discuss the research works addressing each of the challenges chronologically within the respective goals, to highlight the evolution of the proposed solutions over the course of time.

\subsection{Increasing recommendation diversity and novelty}
\label{subsec:increas_nov}

The success of music recommendation algorithms is commonly measured by how well they are capable of making predictions of highly relevant items to a target user, i.e., items that users will likely click, buy, or listen to. Research on MRS has for long solely focused on promoting this perspective through improving \textit{recommendation accuracy}. However, too much personalization, i.e., excessive tailoring to user's existing musical taste can harm the development of users' curiosity. The extreme case of this 
concern is described as the \dquotes{filter bubble} effect, which refers to the scenario where --- as the result of excessive personalization of content geared toward users' individual preference --- personalization services (e.g., web search engines and RS) reduce diversity of the information presented and lead to partial information blindness (i.e., filter bubbles). CF models relying on historical data can be seen as \dquotes{safe} recommendation approaches in most scenarios since their design promotes recommendation of popular items, enabling them to meet an acceptable recommendation accuracy in situations where enough interactions are available. However, these approaches do not allow users to explore and understand their musical taste and increase the risk of trapping users in a self-strengthening cycle of opinions without motivating them to discover alternative genres or perspectives. In other words, CF approaches impede recommendation of novel (or unpopular items). A practical choice to address this issue is to use CBF or hybrid approaches that leverage audio similarity, therefore being insensitive to the \dquotes{popularity bias} described above, and leading to increased item novelty. 

In addition to the standard MRS tasks of rating prediction and recommending a set or list of music items, diversity and novelty aspects are also relevant for automatic playlist generation (APG). 
While the main objective of APG is to create a \textit{coherent} sequence of tracks (cf.~Section~\ref{sec:sequences}), 
these two aspects are often a side goal in the multi-faceted endeavor of creating a great playlist. A possible aim of an APG system is to achieve playlist coherence in some dimension (e.g., rhythm or tempo) while ensuring diversity in another dimension (e.g., variety of artists).
In the following, we discuss research on the related aspects of diversity, novelty, and hubness. Table~\ref{tab:overview_diversity} provides an overview of the reviewed papers.

\setlength{\parindent}{15pt} 
\begin{table*}[t!]
\centering
\caption{\label{tab:overview_diversity}Overview of research works on \textit{increasing recommendation diversity and novelty}.
}
\begin{tabular}{l | c | c | c | c | c | c}
\toprule
\multicolumn{1}{l}{\textbf{\footnotesize{Sub-goal/major method}}}  &\multicolumn{5}{c}{\textbf{\footnotesize{Level of Content}}} \\  \cline{2-6}
\multicolumn{1}{l}{\textbf{\footnotesize{}}}  &\multicolumn{1}{c}{\textbf{\footnotesize{Audio}}} &\multicolumn{1}{c}{\textbf{\footnotesize{EMD}}} &\multicolumn{1}{c}{\textbf{\footnotesize{EGC}}} &\multicolumn{1}{c}{\textbf{\footnotesize{UGC}}} &\multicolumn{1}{c}{\textbf{\footnotesize{DC}}} \\ \hline


\footnotesize{\textbf{Increase diversity}} & & & & & & \\
\footnotesize{\textbullet \ Probabilistic latent fusion~\cite{kazuyoshi_yoshii_2006_1416826}} &\cmark &\cmark & & & & \\
\footnotesize{\textbullet \  Metric learning~\cite{DBLP:journals/taslp/ShaoOWL09}} &\cmark & & & &\\
\footnotesize{\textbullet \  User filtering w.r.t.~diversity and novelty~ \cite{Schedl:2015:TMR:2766462.2767763}} & & &\cmark & \cmark &\\
\footnotesize{\textbullet \  Multi-objective optimization~\cite{DBLP:conf/ismir/OliveiraNMA17}} &\cmark &\cmark &\cmark & &\\
\changed{\footnotesize{\textbullet \  Deep and wide neural network~\cite{DBLP:conf/um/TommaselRG22}}} &\cmark & & & \cmark &\\
\footnotesize{\textbullet \  Impact of personal characteristics on perceived diversity~\cite{Jin:2018:EIT:3209219.3209225}} & \cmark & & \cmark & \cmark & & \\ 
\footnotesize{\textbf{Increase novelty}} & & & & & & \\
\footnotesize{\textbullet \ Graph transformation and optimization~\cite{klaus_seyerlehner_2009_1416972}} & \cmark & & & & & \\
\footnotesize{\textbullet \  Metric learning to rank~\cite{DBLP:journals/taslp/McFeeBL12}} & \cmark & & & \cmark & \\
\footnotesize{\textbullet \  Time-series analysis, stochastic gradient descent~\cite{yajie_hu_2011_1418301}} & & & \cmark & & \\
\footnotesize{\textbullet \ Model of human memory to consider recentness~\cite{DBLP:journals/corr/abs-2003-10699}}& & & & \cmark & & \\
\footnotesize{\textbf{Reduce hubness}} & & & & & & \\
\footnotesize{\textbullet \ Homogenization of GMMs ~\cite{parag_chordia_2008_1415132}} & \cmark & & & & \\
\footnotesize{\textbullet \ Smoothing of GMMs ~\cite{kazuyoshi_yoshii_2009_1415204}} & \cmark & & & & \\ 
\footnotesize{\textbullet \  Item proximity normalization~\cite{schnitzer2012local}} & \cmark & & & & \\\hline
\end{tabular}
\end{table*}

\textbf{Diversity:}
In an early work, \cite{kazuyoshi_yoshii_2006_1416826} propose a hybrid model that combines a rating-based CF, where ratings were collected from Amazon, and an acoustic CBF using polyphonic timbres of the song. For hybridization, the system uses a probabilistic Bayesian network. The study aims to investigate the impact of hybridization on several aspects: 
\textit{(1)} accuracy of recommendation, \textit{(2)} artist diversification, and \textit{(3)} addressing the new-item CS problem. 
Evaluation is performed on audio signals of Japanese songs and the corresponding rating data collected from Amazon. The results show that the hybrid system outperforms both of the constituting CF and CBF approaches in terms of recommendation accuracy and diversity, where the latter is measured by artist variety of the recommended items. Furthermore, both the hybrid system and CBF models can address the new-item problem by recommending a reasonable amount of favorite tracks even in absence of any rating. While the authors propose a principled approach to unify the CF and CBF models, a practical problem is the \textit{computational cost} of computing the parameters of the model in an online manner particularly when new users or items are continuously added to the system/catalog. To address this problem,  in~\cite{DBLP:journals/taslp/YoshiiGKOO08}, the authors extend their previous work and develop an~\textit{incremental learning} method that can cope with increasing numbers of users, allowing new users and items to register to the system incrementally and at low computational cost, without compensating accuracy.

In~\cite{DBLP:journals/taslp/ShaoOWL09}, the authors introduce a hybrid system combining CF and audio-based CBF to improve the diversity of music recommendations\footnote{Note that the authors of \cite{DBLP:journals/taslp/ShaoOWL09} use the term \dquotes{novelty}, but according to the common definition it is diversity what they address.} --- measured in terms of variety of artists --- without compensating the accuracy of recommendation. The proposed system entails a \textit{metric learning} approach, casted in an optimization framework in which the goal is to minimize the distance between two similarity components, one computed based on audio content and the other based  on users' interaction patterns. The system exploits varieties of acoustic features (MFCCs, Daubechies Wavelet Coefficient Histograms (DWCH), spectral centroid, spectral rolloff, spectral flux, ZCR, and low energy rate). This results in designating dynamic weights to different acoustic features from one type of music piece to another type or from one user to another user. A graph representation and label propagation are proposed to leverage music similarities and effect recommendations. Evaluation is performed on a dataset obtained from NewWisdom\footnote{\url{http://www.newwisdom.net}} and shows that improved similarity estimation can enhance the accuracy of recommendations, 
but can also increase diversity. 

In \cite{Schedl:2015:TMR:2766462.2767763}, the authors consider diversity (and novelty) on the user level. They define measures for these aspects (e.g., for diversity the distinct number of fine-grained genres in a user's listening history; for novelty the average share of new items listened to over a shifting time window) and compute respective diversity and novelty scores for each user. Then, a \textit{user-based pre-filtering} strategy is adopted to create different user groups with respect to these scores; and a user-based CF, a CBF approach based on user-generated tags from Last.fm, a location-based CA approach, and combinations of the former three as well as some baselines are evaluated on each group separately.
The authors use a dataset of 200 million listening records from Last.fm and measure performance with respect to precision, recall, and F-measure. They identify significantly different performance depending on the user group under consideration. 
A general finding of their evaluation is that combining CF, CBF, and popularity-based recommendations perform best, in terms of recall and F-measure, for users with a high preference for diversity and users with medium or high preference for novelty.

In~\cite{DBLP:conf/ismir/OliveiraNMA17}, the authors address the diversification problem by studying the accuracy--diversity trade-off casted into a \textit{multi-objective optimization} problem to create a balance between affinity and diversity. The proposed system assesses the impact of multiple aspects including \textit{contemporaneity} (referring to the birth year of the artist or the year the band was formed), \textit{locality} (referring to the artist's birth country), the artist's \textit{gender}, \textit{type} (e.g., solo, band, orchestra), and \textit{genre}. Metadata are collected from Music Brainz\footnote{\url{https://musicbrainz.org}} and DBPedia.\footnote{\url{http://wiki.dbpedia.org}} The effectiveness of the proposed approach is validated on a dataset from Last.fm. 
The authors show that the proposed approach can recommend approximately 2.5 times more artists from different countries, compared to the best baseline.

\changed{
In a recent work, Tommasel et al.~adopt a \textit{deep and wide neural architecture} to increase diversity and novelty of the user's music recommendations~\cite{DBLP:conf/um/TommaselRG22}. While the wide part aims at learning personalized user preferences, the deep part targets generalization to unseen user--item interactions.
At its core, the authors' approach is based on a \textit{knowledge graph} and a \textit{listening graph}, constructed from artists, tracks, (user-generated) tags, Spotify audio features, and user--track interactions, respectively.
A graph convolutional network (GCN) trained on the listening graph is used to create user embeddings that represent each user. Artist, track, and tag embeddings are learned and concatenated with Spotify audio features to represent each track. 
The deep part subsequently concatenates user and track representations and feeds them into a network of 3 dense layers. In contrast, the wide part concatenates the user, artist, track, and tag embeddings (but not the Spotify features) and feeds them into a single dense layer. Ultimately, the predicted user--item interaction score is obtained by adding the predictions of the wide and the deep part.
For training the proposed network, the authors define a loss that favors item diversity by giving higher weights to user--item pairs that are more distant in a different embedding space, where distance is computed as cosine distance on Word2vec user and track embeddings. The authors claim that such items \dquotes{do not belong to the user filter bubble}.
Evaluation against 9 baseline algorithms is carried out on a dataset from Last.fm which the authors acquired themselves.
The proposed approach performed well on diversity and novelty metrics and also competitive in terms of precision and NDCG.}

Another research direction for breaking the filter bubble is to increase users' awareness of their consumption patterns, which has been explored to find blind spots in media consumption and visualize them in movie and music domains~\cite{DBLP:conf/sac/TintarevRS18,Jin:2018:EIT:3209219.3209225}. The authors of~\cite{Jin:2018:EIT:3209219.3209225} investigate the \textit{impact of personal characteristics} such as visual memory (VM) and music sophistication (MS) on the perceived diversity of music recommendations. The intuitive idea behind this work is that music experts with better VM and MS may perceive the same music recommendation list more diverse than non-specialists with lower VM and MS. To test this hypothesis, this work tries to actively engage users in exploring their music consumption profile by proposing two visual interfaces: one capturing the diversity in music recommendation by considering genre and popularity (from Spotify), highlighted by color and size; the other by visualizing audio features in a 2D Cartesian space in a more complex visualization form. Results of a user study with 83 participants show that people with higher MS and VM are likely to perceive the audio feature-based list more diverse than the other list. On the other hand, lower MS results in significantly lower diversity perception for the audio-based list. 

\setlength{\parindent}{15pt} \textbf{Novelty:} 
 {As stated above, CF is prone to promote recommendations of popular (less novel) items. Hybridization of CF and CBF models is the most widely adopted approach to mitigate this problem and improve novelty. In the following, we discuss papers that use various hybridization techniques to approach this goal.}

{One of the earlier works that addresses the question of novelty in recommendations is~\cite{klaus_seyerlehner_2009_1416972}, in which the authors look into the impact of the recommendation network on the \textit{navigability} of MRS results. They define a recommendation network as a graph whose vertices represent music items, and a direct edge between two vertices $v_1 \rightarrow v_2$ means that a recommender that is fed with $v_1$ will recommend $v_2$.
The authors analyze the recommendation network of an audio CBF recommender and show that many items are not recommended because they are not reachable via browsing and remain hidden in the \dquotes{long tail}. The authors propose to construct a \textit{browsing graph} as a transformation of a recommendation graph, via replacing directed edges with undirected ones. Afterwards, edges are discarded in an iterative fashion such that two properties are satisfied: maximum outdegree and minimum indegree. 
The authors show that performing this transformation and optimization, their proposed system can offer substantially more novel item, while maintaining the level of accuracy.

{
In~\cite{DBLP:journals/taslp/McFeeBL12}, the authors propose a \textit{metric learning to rank} (MLR) approach that combines an audio content-based similarity measure --- computed on artist level --- by learning similarities from interaction data. To obtain an audio content representation of items, the system learns a codebook representation of delta-MFCCs to represent each item as a histogram over codewords by using vector quantization. Performing MLR on this representation results in optimizing the feature space such that the system can recommend more novel items based on audio content. 
For evaluation 
McFee et al.~use the CAL10K dataset~\cite{DBLP:conf/mir/TingleKT10} composed of 10,832 songs and show the effectiveness of the compact audio representation and MLR to offer personalized and novel recommendations.} 

Novelty is also addressed in the literature on sequential music recommendation (cf.~Section~\ref{sec:sequences}). 
{For instance, the authors of~\cite{yajie_hu_2011_1418301} aim to recommend the next song in a playlist, by satisfying five aspects: user affinity, novelty/freshness, match with the user's listening pattern, and consistency with genre and recording year profile of the listened songs. 
The latter metadata are gathered from AllMusic.com, an expert-curated music platform.
This work builds on the assumption that recommending the next song based on audio content {similarity} between the target song and the user profile is not an appropriate choice since users differ in their willingness to listen to similar songs or mixing songs according to genre and year. Therefore, the authors propose a \textit{time series analysis} using an autoregressive integrated moving average (ARIMA) to model the user's taste over time and select matching songs to recommend. 
The authors define freshness --- which closely relates to the concept of novelty --- as the strength of strangeness~\cite{ebbinghaus2013memory}, which is regarded as equivalent to the amount of forgotten experience and is captured by a {\dquotes{forgetting curve}}. 
All five aspects mentioned above are combined using a \textit{gradient descent} method, which gives more weight to recently consumed songs, to care for users' needs. 
Evaluation is performed in a user study, using an implementation of the proposed system. Against a baseline that randomly picks songs as the next ones, the authors show that recommendations made by their system result in fewer skips and longer listening duration. }
In a recent work~\cite{DBLP:journals/corr/abs-2003-10699}, the authors adopt a \textit{psychological model of human memory (ACT-R)} to show that there are two factors that are crucial for remembering music: frequency of exposure and recentness of exposure, which strongly relates to novelty in recommendations. In fact, the authors show that a simple MRS that integrates these two aspects outperforms a popularity-based approach, an item-based and a user-based CF, and models that only consider one of the two aspects: recentness or frequency of exposure. 
\setlength{\parindent}{15pt} \textbf{Hubness:} {A natural problem that occurs during music similarity computation is the issue of \dquotes{hubness}, which can significantly decrease the diversity of recommendation lists, thereby hindering the opportunity for music discovery. Hubs are \dquotes{songs that objectively have nothing to do with the reference song}~\cite{Aucouturier:JNRSAS04} and songs which act as a hub --- according to the similarity function used to create recommendations --- remain similar to many other songs, and therefore keep appearing undesirably in many recommendation lists, impeding other songs from being recommended~\cite{arthur_flexer_2012_1417865}. 
The hubness problem particularly affects similarity spaces that are created by applying a similarity metric on high-dimensional data points~\cite{schnitzer2012local}. In the domain of content-based music recommendation, this particularly affects computational features extracted at the audio layer of the onion model, which often results in data spaces of thousands of dimensions, e.g., when flattening the covariance matrix of MFCC summaries~\cite{mandel:mirex:2010,tzanetakis:mirex:2010} or computing block-level features~\cite{seyerlehner2010fusing}.

An early work that addresses the hubness problem is~\cite{parag_chordia_2008_1415132}, in which the authors propose an \textit{ethnical music recommender} for non-Western music by building a CBF recommender that describes each music item with a Gaussian mixture model (GMM) on frame-based MFCC features and uses earth mover's distance to compute similarities. The system addresses the hubness issue by \textit{homogenization of GMMs}, i.e., discarding the degenerate component of GMMs with the goal to obtain more homogeneous (i.e., uniformly distributed) hub histograms.

{Another GMM-based strategy to reduce hubness is presented in~\cite{kazuyoshi_yoshii_2009_1415204}, where the authors propose a probabilistic latent semantic model, extending their previous works~\cite{kazuyoshi_yoshii_2006_1416826,DBLP:journals/taslp/YoshiiGKOO08}. Their model uses GMMs to combine audio features (related to voice) and ratings. 
Yoshii et al.~notice, however, that including both kinds of data creates a {hubness} problem. 
They address this problem by investigating three \textit{GMM smoothing} techniques: \textit{(1)} multinomial smoothing, \textit{(2)} Gaussian parameter tying, and \textit{(3)} artist-based item clustering. Experiments show that the last two approaches significantly help improving recommendation accuracy and reducing hubness. The latter is measured based on the entropy of recommendations.}

Another approach to mitigate the hubness problem is provided in ~\cite{schnitzer2012local}. Schnitzer et al.~propose a normalization strategy called \textit{mutual proximity}, which corrects distortions in high-dimensional similarity spaces that would otherwise result in the creation of hubs. Their idea is to only consider an item $i$ as nearest neighbor of another item $j$ if $j$ also appears among the neighbors of $i$. They formulate this idea in a probabilistic model and show that applying this model on a large variety of datasets indeed reduces hubness; they measure hubness as the skewness of the distribution of k-occurrences across all items (i.e., the frequency with which an item occurs among the k nearest neighbors of all other items in the recommendation list).

\subsection{Providing transparency and explanations}
\begin{table*}[!tb] 
\centering
\caption{\label{tab:overview_transparency}Overview of research works on \textit{providing transparency and explanation}.}
\begin{tabular}{l | c | c | c | c | c | c}
\toprule
\multicolumn{1}{l}{\textbf{\footnotesize{Sub-goal/major method}}}  &\multicolumn{5}{c}{\textbf{\footnotesize{Level of Content}}} \\  \cline{2-6}
\multicolumn{1}{l}{\textbf{\footnotesize{}}}  &\multicolumn{1}{c}{\textbf{\footnotesize{Audio}}} &\multicolumn{1}{c}{\textbf{\footnotesize{EMD}}} &\multicolumn{1}{c}{\textbf{\footnotesize{EGC}}} &\multicolumn{1}{c}{\textbf{\footnotesize{UGC}}} &\multicolumn{1}{c}{\textbf{\footnotesize{DC}}} \\ \hline

\footnotesize{\textbf{Neighborhood-based explanations}} & & & & & & \\
\footnotesize{\textbullet \ Visualizing ratings of target user's neighbors~\cite{DBLP:conf/cscw/HerlockerKR00}
} & & & & \cmark & \\

\footnotesize{\textbf{Feature-based explanations}} & & & & & & \\
\footnotesize{\textbullet \ 
Descriptive terms extracted from web pages~\cite{DBLP:conf/ismir/PampalkG07}} & \cmark & & \cmark & \cmark & \\
\footnotesize{\textbullet \ Overlap and difference tag clouds~\cite{green_etal:recsys:2009}} & & & \cmark & \cmark & \\

\footnotesize{\textbf{Contextual explanations}} & & & & & & \\
\footnotesize{\textbullet \ Create an affective artist space to navigate~\cite{Andjelkovic:2016:MIM:2930238.2930280}} & \cmark & & \cmark & & \\
\footnotesize{\textbullet \ Reinforcement learning~\cite{10.1145/3240323.3240354}} & & \cmark & \cmark & \cmark & \\

\footnotesize{\textbf{Audible explanations}} & & & & & & \\
\footnotesize{\textbullet \ Story generation to explain song transitions~\cite{DBLP:conf/ismir/BehroozMTKC19}} & & \cmark & \cmark & & \\
\footnotesize{\textbullet \ AudioLIME for segmentation and source separation~\cite{DBLP:conf/ecir/MelchiorreHSW21}} & \cmark & & & & & \\

\changed{\footnotesize{\textbf{User characteristics-based explanations}}}   & & & & & & \\
\footnotesize{\textbullet \ \changed{Contrastive learning for explaining negative preferences~\cite{DBLP:conf/recsys/ParkL22}}
} & \cmark & & & & \\
\footnotesize{\textbullet \ \changed{Tailor explanations to personal characteristics~\cite{DBLP:journals/umuai/MillecampCV22}}
} & \cmark & & & & \\
\hline
\end{tabular}
\end{table*}

The recent and ongoing discussion about ethics (e.g., trustworthiness and credibility) of machine learning algorithms has stimulated much research on these aspects, also in the context of RSs in general, and MRSs in particular.
While we notice a certain level of vagueness and ambiguity when it comes to an exact terminology of such aspects, most often they can be categorized into \textit{transparency} and \textit{explainability}. The former typically refers to unveiling the exact mode of operation and internal mechanisms of the recommendation algorithm or model, whereas the latter is concerned with creating (and presenting to the user) explanations of why the items in the recommendation list have been recommended.
This offers the user a means to understand the reasoning of the system when recommending a certain item or list of items.
Several studies have revealed that both aspects (transparency and explainability) can increase user engagement and trust in the RS~\cite{Tintarev2015,DBLP:journals/aim/FriedrichZ11,Andjelkovic:2016:MIM:2930238.2930280,ribeiro_etal:kdd:2016,DBLP:conf/sigir/BalogR20}.

Zhang et al.~\cite{DBLP:journals/ftir/ZhangC20} provides a categorization of explanation types for RSs in a domain-independent way. An in-depth survey and position paper on explainability of MRSs can be found in~\cite{afchar:aimag:2021}.
Unifying and extending the findings of these seminal works, we categorize the reviewed articles into \textit{neighborhood-based explanations}, 
\textit{feature-based explanations}, 
\textit{contextual explanations}, and \textit{audible explanations}.
\changed{Recently, some works that directly incorporate specific user characteristics in the process of creating explanations have emerged. We, therefore, include another category \textit{user characteristics-based explanations} in our survey.} Table~\ref{tab:overview_transparency} shows an overview of the discussed literature.

\setlength{\parindent}{15pt} \textbf{Neighborbood-based 
explanations:} 
This category of explanations comprises the simplest, and historically earliest, kind of techniques. Directly derived from user- or item-based CF approaches, explanations use only the similarities between users or between items the target user listened to. Such explanations, therefore, commonly are of the kind \dquotes{because other similar users liked the recommended item} or \dquotes{because the recommended item is similar in terms of other users' consumption/rating behavior}.
In the music domain, which historically has a focus on content-based approaches, this kind of explanation is very rare. Notwithstanding, domain-independent examples are given in \cite{DBLP:conf/cscw/HerlockerKR00}, where the authors propose to present to the target user a visualization of his or her neighbors' (most similar users) rating distribution on the recommended item.

\setlength{\parindent}{15pt} \textbf{Feature-based explanations:}
This explanation type is tied to content-based recommendation approaches. In particular, content features or descriptors from different levels of our onion model are used to create explanations such as \dquotes{because you like complex melodies} (audio), \dquotes{because the song's genre is one of your favorite genres} (EMD), or \dquotes{because it is a cover version of a song you love} (DC).

Early works belonging to this category typically incorporate expert- or user-generated descriptors into a music exploration and recommendation interface. For instance, Pampalk et al.~\cite{DBLP:conf/ismir/PampalkG07} present the \textit{MusicSun} system, which extracts descriptive terms from web pages about music artists to characterize them, using predefined vocabularies of genres, instruments, moods, and countries.
Recommendations are created for one or more seed artists the user selects, based on a weighted combination of audio similarity, similarity of TF-IDF vectorized artist web pages, and tag similarity computed from the seed artists' descriptive terms. Each recommended artist is shown with accompanying terms for an explanation.
The user can then refine or update the recommendations by selecting one of the nine most representative terms for the set of seed artists. Therefore, the same vocabulary is used to steer the recommendation process and to explain recommendations.

Stephen et al.~\cite{green_etal:recsys:2009} present a tag-based explanation approach for artist recommendations that leverages user-generated tags extracted from Last.fm and expert-generated Wikipedia descriptions to create an \dquotes{aura of musical artists}, which is essentially a tag cloud.
The authors propose a system that addresses the simple recommendation scenario of requesting similar artists to a given seed artist. Similarity is computed based on the standard TF-IDF vector space model and computing cosine similarity between the vector representations of artists.
Explanations are then created in two ways, an \textit{overlap tag cloud} and a \textit{difference tag cloud}, to visualize what the artists have in common and what distinguishes the recommended artist from the seed artist, respectively.
The overlap tag cloud is created by visualizing the intersection of tags (of the seed artist and the recommended artist); the difference tag cloud as the set difference of the recommended artist with respect to the seed.

\setlength{\parindent}{15pt} \textbf{Contextual explanations:}
This kind of approaches consider contextual factors of music listening to generate explanations, such as mood~\citep{Andjelkovic:2016:MIM:2930238.2930280} or time~\citep{10.1145/3240323.3240354}.

Andjelkovic et al.~\cite{Andjelkovic:2016:MIM:2930238.2930280} create a latent affective space based on artists' mood tags, which is visualized on a 2-D plane using correspondence analysis. This visualization is part of the proposed \textit{MoodPlay} interface for artist exploration and recommendation. Mood tags are acquired from the Rovi platform\footnote{\url{http://developer.rovicorp.com}} and audio content features describing timbre, tempo, loudness, and key, according to~\cite{DBLP:conf/ismir/McFeeL11}. \textit{MoodPlay} creates recommendations based on one or more user-selected seed artists, integrating a mood-based and an audio-based similarity measure.
Defined by the mapping of the seed artist(s) to the mood space, an avatar is positioned accordingly on the mood plane. The user can move the avatar to navigate through the mood space, thereby exploring the music collection in terms of audio and mood similarity, using the mood terms for both interaction and explanation.

McInerney et al.~\cite{10.1145/3240323.3240354} propose an approach called BAndits for Recsplanations as Treatments (\textit{BART}). The authors apply reinforcement learning to model the interactions between items and corresponding explanations, conditioned on both user and item contextual characteristics. For this purpose, the authors adapt a multi-armed bandit approach with a reward function that reflects the user engagement as the result of the explanation in a given context.
Contexts include listening time (\dquotes{because it's Friday}), item novelty (\dquotes{because it's a new album}), genre (\dquotes{because you like Viking Metal}), and item popularity (\dquotes{because it's popular}).
Results of an offline evaluation and of A/B testing confirm that personalizing the explanations in terms of context factors is key to improve user satisfaction, approximated as NDCG and stream rate, i.e., the number of times a user streamed at least one track from a recommendation list.

\setlength{\parindent}{15pt} \textbf{Audible explanations:} 
A different and recent strand of research aims at providing explanations in perhaps the most natural form when it comes to MRS, i.e., explanations the user can actually listen to.
These can take the form of voice explanations similar to a moderator on a radio station~\cite{DBLP:conf/ismir/BehroozMTKC19} or music explanations that refer to the most relevant audio segments of a song~\cite{DBLP:conf/ecir/MelchiorreHSW21}.

The approach proposed in Behrooz et al.~\cite{DBLP:conf/ismir/BehroozMTKC19} tackles the problem of generating explanations, in particular in APG and APC settings. Based on a music playlist, their prototype system automatically creates voice explanations to moderate the transitions between consecutive songs, using story telling techniques. For this purpose, editorial metadata (artists, albums, songs, genres) and facts extracted from Wikipedia are leveraged. Individual explanations for transitioning between one pair of consecutive tracks are created by inserting these pieces of information into natural language templates.
Creating explanations for all transitions in a given playlist is achieved by modeling the task as a graph traversal problem over the space of all possible stories, optimizing a weighting function that considers, for instance, repetitions, lengthiness of explanations, or positional preferences.
A qualitative evaluation of the proposed system by semi-structured interviews was carried out.

In a recent work, Melchiorre et al.~\cite{DBLP:conf/ecir/MelchiorreHSW21}~adopt \textit{AudioLIME}~\citep{DBLP:journals/corr/abs-2008-00582} to create listenable explanations of content-based music recommendations. The proposed system, called \textit{LEMONS}, creates recommendations based on a \changed{convolutional neural network (CNN)} trained on chunks of audio; positive samples are drawn from the user's listening history, negative ones randomly sampled.
Thereafter, AudioLIME computes interpretable audio components using source separation and temporal segmentation techniques. These components are fit to the output of the MRS model using linear regression, thereby assigning a weight to each time--source pair of the recommended music pieces. The audio of pairs with the highest weights is offered as explanation to the user.

\changed{\textbf{User characteristics-based explanations:} This kind of explanations directly use characteristics of users, such as preferences~\cite{DBLP:conf/recsys/ParkL22} or psychological traits~\cite{DBLP:journals/umuai/MillecampCV22}, to make the recommendation algorithm more transparent for the user.}

\changed{For example, Park and Lee~\cite{DBLP:conf/recsys/ParkL22} use contrastive learning to create a MRS that not only learns from positive feedback (songs the user likes), but also from negative feedback (songs disliked by the user). 
The proposed contrastive learning exploiting preference (CLEP) models are based on Siamese neural networks (SNNs)~\cite{koch2015siamese}.
The authors design a content-based MRS leveraging audio features which are obtained from pre-trained self-supervised models.
To acquire the user's preference indications, including liking \textit{and} disliking information on songs, Park and Lee perform a web-based user study.
The embedding space is then learned in a way such that either positive preferences (CLEP-P), negative preferences (CLEP-N), or positive and negative preferences (CLEP-PN) are enforced. Their experimental results show that CLEP-N yields the highest accuracy and precision, at the same time by far the lowest false positive rate, among the three models. The authors conclude that leveraging negative preferences has great potential for explaining recommendations because knowing what users do not want to listen to could be as informative for them as knowing what they like.
}

\changed{Millecamp et al.~\cite{DBLP:journals/umuai/MillecampCV22} investigate which kind of explanation is best suited for different user groups, where groups are defined by various personal characteristics, in particular, the target user's need for cognition (i.e., ``the tendency to engage in and enjoy effortful cognitive
activities''~\cite{doi:10.1207/s15327752jpa4803_13}), musical sophistication (music expertise of users)~\cite{10.1371/journal.pone.0089642}, and the personality trait of openness~\cite{https://doi.org/10.1111/j.1467-6494.1992.tb00970.x}.
Users are categorized into high and low levels of each of these three characteristics. The authors conduct a user study and eventually formulate, based on its results, some guidelines on which explanations should be presented and how they should be tailored to the investigated personal characteristics of users. 
The explanation types Millecamp et al.~looked into are: (1) bar chart displaying the differences between the (Spotify) audio features of the recommended song and the user's preferred songs, for each dimension (popularity, happiness, energy, and danceability), 
(2) the percentage of the match between the recommended song and the overall user preference, 
(3) scatter plot of the recommended songs over the dimensions of popularity, happiness, energy, and danceability, and 
(4) information about the song from the user's playlist that was used to create the recommendation.
The authors find that low-sophistication users prefer simple over complex explanations, that both high- and low-need-for-cognition users prefer being given all explanations at once, and that low-openness users prefer a smaller number of explanation styles (in particular, bar charts).
}

\subsection{Accomplishing context-awareness} \begin{table*}[!t]
\centering
\caption{\label{tab:overview_context-awareness}Overview of research works on \textit{accomplishing context-awareness}.}
\begin{tabular}{l | c | c | c | c | c | c}
\toprule
\multicolumn{1}{l}{\textbf{\footnotesize{Sub-goal/major method}}}  &\multicolumn{5}{c}{\textbf{\footnotesize{Level of Content}}} \\  \cline{2-6}
\multicolumn{1}{l}{\textbf{\footnotesize{}}}  &\multicolumn{1}{c}{\textbf{\footnotesize{Audio}}} &\multicolumn{1}{c}{\textbf{\footnotesize{EMD}}} &\multicolumn{1}{c}{\textbf{\footnotesize{EGC}}} &\multicolumn{1}{c}{\textbf{\footnotesize{UGC}}} &\multicolumn{1}{c}{\textbf{\footnotesize{DC}}} \\ \hline

\footnotesize{\textbf{Leverage spatial context}} & & & & & & \\
\footnotesize{\textbullet \ Fusion of auto-tagging and knowledge graph 
~\cite{Kaminskas:2013:LMR:2507157.2507180}} & \cmark &  & \cmark & \cmark & & \\
\footnotesize{\textbullet \  
Mapping music and locations into joint space~\cite{cheng2016effective}} &\cmark &\cmark &\cmark & &\\
\footnotesize{\textbullet \ Hybrid memory-based with GPS data ~\cite{Schedl:2014:LMA:2695416.2695439}} & \cmark & & \cmark & \cmark &\\ 
\footnotesize{\textbullet \ Integration of many (location-based) web services ~\cite{DBLP:journals/percom/AlvarezZB20}} & \cmark & \cmark & \cmark & \cmark &\\ 
\footnotesize{\textbullet \ \changed{Concentration level estimation to recommend music while working~\cite{DBLP:journals/umuai/YakuraNG22}}} & \cmark & & & & \\ 


\footnotesize{\textbf{Leverage affective context}} & & & & & & \\
\footnotesize{\textbullet \  Map film music features to emotion~\cite{Kuo:2005:EMR:1101149.1101263}}  & \cmark & & \cmark & & & \\
\footnotesize{\textbullet \  Reasoning in ontology relating music to emotion~\cite{Rho:2009:SMM:1631272.1631395}} & \cmark & \cmark & \cmark & & \\
\footnotesize{\textbullet \  Include mood into factorization machine~\cite{Chen:2013:UEC:2502081.2502170}} & \cmark & \cmark & & \cmark & \\
\footnotesize{\textbullet \ Memory-based CF with emotional similarity~\cite{deng2015exploring}} & & & & \cmark & \\
\footnotesize{\textbullet \ \changed{Audio-based mood classification with CNNs~\cite{DBLP:conf/recsys/BontempelliCRML22}}} & \cmark & & \cmark & & \\


\footnotesize{\textbf{Leverage social context}} & & & & & & \\ 
\footnotesize{\textbullet \ Rating diffusion through online social network~\cite{mesnage_etal:womrad:2011}} & & \cmark & & \cmark & & \\
\footnotesize{\textbullet \ Integrate social influence into user-based CF~\cite{DBLP:conf/dcai/Sanchez-MorenoP18}}  & & \cmark & & \cmark & \\
\footnotesize{\textbullet \ Factor graphical model of social influence~\cite{DBLP:journals/mta/ChenYZ19}}  & & \cmark & & \cmark & \\ 
\footnotesize{\textbullet \ Audio, cultural, and socio-economic user model~\cite{zangerle:tismir:2020}} & \cmark & \cmark & & \cmark & & \\
\footnotesize{\textbullet \ Music-cultural country clusters and VAE ~\cite{10.3389/frai.2020.508725}} & & \cmark & & \cmark & \\ 
\hline
\end{tabular}
\end{table*}

Considering situational or contextual aspects of the user when recommending music items has been identified as an important factor to increase user experience, e.g.\changed{~\cite{10.1145/3320435.3320445,Schedl2018,lozano2021context}}.
Such contextual information can be categorized in manifold ways, cf.~\cite{bauer_novotny:jaise:2017} for taxonomies of context factors. As one of the earlier works, \cite{schilit1994context} classifies context based on: \textit{computing environment} (available processors, network capacity),\textit{ user environment} (location, social situation), and \textit{physical environment} (lighting and noise level). 
\cite{abowd1999towards} categorizes context into \textit{primary level} which can be measured directly (location, identity, activity, and time) and \textit{secondary level} (such as emotional state of the user). \cite{aggarwal2016context} classifies context into 
\textit{time}, \textit{location} and \textit{social information} (user's friends, tags, and social circles). Finally, \cite{kaminskas2012contextual} provides a categorization of 
contextual variables into: \textit{item-related} (location, time, weather, other parameters such as traffic, noise level or traffic jam), \textit{user-related} (activity, demographic and emotional state), and \textit{multimedia context} (text, image, music, video). 

We acknowledge the diversity of these different categories and taxonomies of context aspects.
For the purpose of the survey at hand, we adopt a pragmatic perspective and organize our discussion according to the types of context predominantly used in relevant work on MRS into \textit{spatial context} (e.g., GPS coordinates, location, place of interest), \textit{affective context} (e.g., mood or emotion), and \textit{social context} (e.g., friends or cultural background).

\setlength{\parindent}{15pt} \textbf{Spatial context-awareness:}
Research works leveraging spatial information can be roughly categorized into those using specific places of interest (e.g.,~\cite{Kaminskas:2013:LMR:2507157.2507180}), those using generic locations such as park, gym, or theater (e.g.,~\cite{cheng2016effective}), and those using exact location data, commonly from GPS data (e.g.,~\cite{Schedl:2014:LMA:2695416.2695439}).

Belonging to the first category, \cite{Kaminskas:2013:LMR:2507157.2507180} propose a context-aware MRS that suggests music that matches a given place of interest (POI), such as a monument, a church, or a park. This matching is achieved by 
\textit{(1)} connecting items between both domains (music pieces and POIs) via tags assigned by users, chosen from a single tag vocabulary, reflecting their emotion towards these items; \textit{(2)} building a knowledge-based RS using DBpedia.\footnote{\url{https://wiki.dbpedia.org}} The choice of emotional tags leveraged for (1) is motivated by the fact that both sets of items (music pieces and POIs) can evoke strong emotions, and the commonalities can be leveraged to estimate a degree of matching between them~\cite{braunhofer2013location}.
The final music recommendation approach is a hybrid combination of an \textit{auto-tagging-based} matching created from (1) and the \textit{knowledge-based} approach (2).
The auto-tagger uses MFCCs and block-level spectral features~\cite{seyerlehner2010fusing} from the audio to predict a set of emotion tags for each music item. It is trained from human annotations of a subset of the used music collection. The Jaccard index between the set of emotion tags of a given POI (also human-created) and each music item's tags is then used as similarity measure to make recommendations.
The second, knowledge-based, approach establishes a relationship between music-related and POI-related entities and properties, and estimates a matching score from the strength of these relationships.
Given a POI, the recommendations of these two components, i.e.,~matching via auto-tagging and via knowledge-graph, are combined using Borda rank aggregation, to create the ultimate list of recommendations.
Evaluation of the hybrid system (and its individual components) is performed in an online user study, where users judged the appropriateness of the music recommended for 25 POIs. The results show that simple personalization (e.g., via a user's genre preferences) is not sufficient for the task and that the combination of complementary sources (user-generated tags and a knowledge graph) outperforms single-source approaches.
A limitation of this work is related to the dataset, in particular, the limited number and high specificity of the POIs, e.g., the Eiffel Tower in Paris or La Scala in Milan. Therefore, results may not generalize to other POIs, e.g., other towers or opera houses, respectively. Also, the tags for the POIs were elicited in an artificial setting, without the user actually being at the corresponding POI.
As an alternative, music listening behavior could be monitored together with precise location information from smart phones~\cite{deng2015exploring}.

Addressing these limitations,~\cite{cheng2016effective} propose a venue-aware MRS where the venue is defined as the place in which an activity or event takes place, e.g., library, gym, office, or mall. 
The key difference to \cite{Kaminskas:2013:LMR:2507157.2507180} is the way in which the links between location and music are established. While both works exploit audio content to represent music, 
\cite{Kaminskas:2013:LMR:2507157.2507180} use tags learned from acoustic features and user's location to find the association whereas \cite{cheng2016effective} map the characteristics of venues and music into a \textit{latent semantic space}, where the similarity between a music item and a location 
can be directly measured. 
To this end, \cite{cheng2016effective} 
propose a system composed of two main modules: a music concept sequence generator (MCSG) and a location-aware topic model (LTM). The role of the MSCG is to map a music playlist into a sequence of semantic concepts, while the LTM then represents songs and venue types in a shared latent space. 
More precisely, first the audio stream of the given playlist is segmented into shorter units from which four types of acoustic features are extracted: timbral, spectral, tonal, and temporal. Subsequently, based on these audio features and a concept-labeled music collection (concepts are genre, mood, and instrument), the MSCG module is trained to later predict a concept for each audio segment.
The LTM component then learns a latent topic space in which both songs and venues are represented as topic vectors, i.e., probabilistic distributions over the latent topics. Newly added songs are automatically converted into a set of latent space variables using MCSG and LTM.
The relevance of a song for a venue is measured as Kullback-Leibler divergence.
Evaluation is performed on two datasets: TC1 and TC2. The former 
contains songs for eight different venue types (e.g., \textit{library}, \textit{gym}, and \textit{bedroom}), collected from Grooveshark and annotated by nine subjects regarding their fit to each venue type.\footnote{\url{https://groovesharks.org}} Each venue is assigned on average 187 songs.
 
TC2 is a collection of 10,000 highly popular songs from Last.fm, for which the authors obtained the corresponding audio from 7digital.\footnote{\url{https://www.7digital.com}} \cite{cheng2016effective} show that their approach of representing songs and venues in a shared latent space outperforms using only audio features, for most venue types, in terms of (average) precision and NDCG metrics.
A limitation of this approach is the small number of considered venues.
Unlike the above-mentioned works, which leverage semantic information about locations (either POIs or categories of locations/venues), other approaches use more precise positional information, commonly acquired through GPS signals.
For instance, \cite{Schedl:2014:LMA:2695416.2695439} propose hybrid MRS approaches that combine similarity information derived from the songs' audio signal, from artists' web pages, and from users' GPS data shared on Twitter.
The proposed systems take a set of songs, artists, and locations each user has listened to at a certain place as input. From the songs, several timbral (e.g., MFCCs), rythmic (e.g., onset patterns), and some highly specific audio features (e.g., ``attackness'') are extracted.
From the top web pages about the artists, found by a search engine, TF-IDF features are computed. Subsequently, the song-level audio features are aggregated to the artist level, and a joint similarity measure incorporating the audio and web features is computed as a linear combination of both. To rectify the resulting similarity space and avoid hubs, Schedl et al.~covert the similarities to mutual proximities~\cite{schnitzer2012local}.
Location-based distance between two users is defined as distance between their listening events' centroids (using linear or exponential weighting). 
Leveraging these artist similarities and user distances, the authors adopt \textit{memory-based top-N approaches} to recommend artists. They use individual recommenders for CBF (on artist similarities), for CF (on user--item interactions), and for a location-aware recommender (basically a CF uses location-based user distances to identify the active user's neighbors).
The final recommendation list is assembled as the union of the individual recommenders' suggestions.
Evaluation is carried out using the MusicMicro dataset\footnote{\url{http://www.cp.jku.at/datasets/musicmicro}} of geolocated tweets~\cite{10.1007/978-3-642-36973-5_87}, containing more than 500K listening events created by more than 100K users.
Results indicate that hybrid MRS that include location information have higher accuracy than approaches using only one data source.
Potential limitations include the simple memory-based recommendation approaches. It can be assumed that more sophisticated methods yield higher performance. 

A recent work that also exploits GPS data is \cite{DBLP:journals/percom/AlvarezZB20}. The authors propose an application for runners that integrates many \textit{web services}, including location-based services.
They incorporate, among others, audio features and location information into the recommendation pipeline. Audio features are acquired from Spotify.\footnote{\url{https://developer.spotify.com/web-api/get-several-audio-features}}
Leveraging GPS data, the system considers the geographic area the runner typically runs (e.g., parks or forests), the altitude, and the type of soil (e.g., sand or gravel).
The recommendation algorithm at its core is a \textit{top-N nearest neighbor approach extended by several filters}.
One component is a rule learner that infers the runner's emotion from the location attributes described above. Feeding the predicted current emotion of the target user into a number of trained regression models, a vector of Spotify audio features suited for the user when exhibiting a certain emotion is predicted. This vector is then used to identify the nearest song neighbors in Spotify's music catalog, adopting the efficient Annoy algorithm for approximate nearest neighbor search.\footnote{\url{https://github.com/spotify/annoy}}
The resulting list of song candidates is subsequently filtered by \textit{(1)} the target user's preferred genres elicited at sign-up to the app and \textit{(2)} a memory-based CF component that is based on the cosine distance between the user--genre matrix.
\'Alvarez et al.~evaluate their application with 500 runners in the Spanish city of Zaragoza. Surprisingly, the evaluation focuses on service response times rather than relevance and utility of recommendations, which is arguably a limitation of their work.
Another limitation is the paper's strong focus on implementation details and the use of existing web services. At the same time, it lacks details about the scientific methodology, e.g., about the adopted regression models. Also, the quite uncommon choice of leveraging user--genre interactions instead of user--track or user--artist interactions is not motivated well.

\changed{The work by Yakura et al.~\cite{DBLP:journals/umuai/YakuraNG22} is a novel and intriguing study since it pays attention to the role that music plays in the workplace (and considering the concentration level of the user). The work is based on the assumption that songs that the user likes very much and songs the user dislikes very much might have negative effects on concentration; thus, this information should be considered when designing a MRS for consumption while working. To locate good music, authors use audio characteristics. Specifically, they assess the mean and variance of average MFCC values across the entire song, 
local spectral qualities (centroid, roll-off, flux, and zero-crossings) across the entire song, 
and the chorus pace. 
PCA is applied for dimensionality reduction. Experimental validation underlines the validity and accuracy of the suggested technique, including concentration estimation. In addition, a user study confirms the appropriateness of the recommendation results based on the participants' observed behavior and comments. A number of limitations of this work include the following: prior to evaluating the effect based on observable data, it is preferable to measure the concentration level of participants using the systems and analyze how songs affect their concentration level to validate the usefulness of the suggested strategy (e.g., the number of operations). The use of psycho-physiological sensors is a viable option, however, such data may not adequately reflect the internal state of participants \cite{zuger2015interruptibility}. Second, the  number of participants was relatively limited to ensure the generalizability of the findings. In addition, the accuracy of the concentration estimation might be altered by the experimental settings, as it is highly dependent on the nature of the task.} 

\setlength{\parindent}{15pt} \textbf{Affective context-awareness:}
Emotion and mood have been shown to significantly influence human behavior and preferences, among others for music~\cite{Rentfrow2011MusicFFM,eerola_vuoskoski:2011,Ferwerda:2017:PTM:3079628.3079693}, making these psychological constructs valuable for context-aware MRS.
Emotion is an affective response of high intensity to a particular stimulus, while mood refers to an affective state of longer duration albeit lower intensity.\footnote{Please note that while psychological research clearly distinguishes between emotion and mood, RS research often uses these terms interchangeably.}
Approaches to affect-aware MRS are often tied to the task of music emotion recognition (MER) \cite{DBLP:journals/tist/YangC12} that aims at predicting emotion from the audio signal of a music piece, in order to establish a connection between the user's affective state and the item's affective annotation. 

In an early work on the topic, Kuo et al.~\cite{Kuo:2005:EMR:1101149.1101263} propose an emotion-based MRS model that leverages 
the \textit{relationship between audio features and emotions perceived in film music}. The authors extract melodic, rhythmic, and tempo features from the MIDI representation of film music 
and create an \textit{affinity graph} to model associations between these audio features and emotions manually assigned to each music track. 
The proposed recommendation model takes a user's emotion, expressed as a query, and returns a set of audio features created by a random walk model through traversal of the affinity graph.
These features are then used to rank all items in the music collection and provide recommendations accordingly.
Evaluation is carried out as an emotion prediction task with a self-assembled and annotated collection of 107 film musics from 20 animated movies. Experimental results show that the proposed approach can identify the emotion of a music track with an accuracy of up to 85\%.
A limitation of Kuo et al.'s approach is its focus on emotion recognition rather than on the recommendation model, which in fact only considers a simple ad-hoc retrieval setting.

Also based on MER, Rho et al.~\cite{Rho:2009:SMM:1631272.1631395} propose a mood-aware MRS that derives \textit{similarity between music tracks from emotions predicted from tracks' audio features}. These similarities are then used in a CBF and an \textit{ontology-based system}. 
More precisely, a set of audio features describing acoustic properties (e.g.,~tonality, key, energy, tempo, and harmonics) is extracted and a support vector regression (SVR) model is learned to predict the mood of a song on the dimensional valence--arousal plane.
This results in an emotion vector, according to Thayer's model~\citep{thayer1990biopsychology}, assigned to each song. 
The similarity between pairs of songs is subsequently computed as Pearson correlation coefficient.
The authors propose two recommendation approaches: \textit{(1)} an item--item CBF\footnote{The authors refer to this approach as collaborative filtering. However, from their (vague) description, it appears to be a CBF recommender that does not incorporate other users' listening histories.} and \textit{(2)} an ontology-based approach, where the latter infers the target user's mood from contextual data such as time and location. Recommendations are then created via \textit{rule-based and ontology-based reasoning}.
Evaluation experiments are conducted on 41 pop songs. The authors create their own ontology, based on the Music Ontology~\cite{raimond_etal:ismir:2007} and MusicBrainz data.
The SVR-based mood predictor achieves an accuracy of about 88\%.
Performance of the recommenders is only evaluated in terms of reasoning time, which is a limitation of Rho et al.'s work. Others include the lack of quite a few methodological details, particularly the recommendation part, and the rather small dataset used for evaluation. 

\changed{The music streaming service Deezer is the focus of the paper by~\cite{DBLP:conf/recsys/BontempelliCRML22}, which proposes \textit{Mood Flows}, a service to create personalized playlists using CF and filtering songs by moods. This work examines six moods: Chill, Focus, Melancholy, Motivation, Party, and You and Me. Mood information is gathered through (i) the use of mood annotations from music experts and (ii) large-scale mood classification utilizing audio content. 
The music experts manually annotated thousands of songs that, according to their knowledge, comply or do not fit with each of the six predetermined moods. The latter approach is motivated by the impossibility of music curators manually annotating all the recommendable content in the catalog, which contains millions of songs. To address this issue, the authors leverage audio content. Specifically, they calculate a 256-dimensional embedding vector for each song based on its audio input using a VGG-like CNN developed for music audio tagging tasks~\cite{choi2016automatic}. This model is accessible to the public using the musicnn Python library~\cite{pons2019musicnn}. These vectors were then utilized as input for six random forest binary classifiers. Through the use of CF information, audio content analysis, and mood information, Flow Moods is able to build large-scale, personalized playlists tailored to each listener's individual mood. The suggested system for mood classification relies exclusively on auditory signals, which are complemented by lyrics data, which is a limitation. It was conceivable to expand the existing system to deliver additional contextual recommendations, such as a "Christmas" version of Flow in December.} 

Several works approach the challenge of building an emotion-aware MRS by extracting emotions from user-generated texts, in particular from  microblogs~\cite{Chen:2013:UEC:2502081.2502170,deng2015exploring,Cai:2007:MCM:1291233.1291369}. 
This research is based on the assumption that users'  
general moods or emotions felt in response to listening to a music piece are reflected in their writing. 
As an example, Chen et  al.~\cite{Chen:2013:UEC:2502081.2502170} propose an emotion-aware system that suggests music pieces to a user based on emotions extracted from texts shared on the social blogging website LiveJournal.\footnote{\url{https://www.livejournal.com}}
Music recommendations are created using a \textit{factorization machine} (FM) trained on a variety of audio, textual, and mood features next to listening data.\footnote{LiveJournal allows the authors to add mood and music labels to their shared articles.}
In detail, Chen et al.~gather audio features from The EchoNest,\footnote{Formerly \url{http://www.echonest.com}, now owned by Spotify.}  including loudness, mode, and tempo. They compute TF-IDF features from the user-generated articles and further use the affective norms for English words (ANEW)~\cite{bradley1999affective} lexicon to derive valence, arousal, and dominance scores that characterize the user's mood expressed in their writing. In addition, the mood labels directly assigned by users to their texts --- a feature of LiveJournal --- are considered.
The FM model then incorporates these features as interaction parameters between users and items to make context-aware music recommendations.
Evaluation is carried out on a dataset gathered from LiveJournal, including 20K users who created 226K articles and listened to 30K songs. The authors compare their proposed FM-based approach to several CF variants and show that including user and song metadata, audio features, TF-IDF, and mood features altogether yields the best performance in terms of mean average precision (MAP) and recall. A possible limitation of Chen et al.'s work is their use of functionality and features which are specific to the used data source, LiveJournal. To which extent their approach generalizes to other data sources, therefore, remains unclear.

Deng et al.~\cite{deng2015exploring} also leverage user-generated texts to infer mood and listening behavior, and subsequently use this information to build an MRS.
For this purpose, the authors adopt a lexicon-based approach 
to infer emotion categories (up to 21) from texts shared on the Chinese microblogging service Sina Weibo.\footnote{\url{https://www.weibo.com}}
They connect these inferred emotions to the music piece posted directly before the user's sharing of a music listening event. 
A user's emotion is represented as an emotion vector (up to 21-dimensional) holding the term frequencies of the corresponding words in each emotion category. This results in triplets of user, song, and emotion vector.
To create music recommendations, Deng et al.~investigate a user-based and an item-based CF approach, a hybrid, and a graph-based model.
The CF models integrate emotional similarity (cosine of the emotion vectors) between users or between items as additional weighting term into a memory-based approach.
The hybrid variant linearly combines the predicted scores of the user-based and the item-based CF.
The graph-based approach models the user--song--emotion triplets as a bipartite graph and adopts \textit{personalized PageRank} to create recommendations.
Evaluation is conducted on a dataset crawled from Sina Weibo, containing about 29K users, 59K songs, and 1M user--song-emotion triplets.
Results show that user-based CF outperforms the other variants (even the hybrid) in terms of precision, recall, and F1 score. A limitation of this work is the simple memory-based recommendation approaches that the authors adopt.

\setlength{\parindent}{15pt} \textbf{Social context-awareness:}
Works in this category leverage either data from social networks (e.g., users' friendship connections) or data about the cultural background of the users. Note that there is some overlap between social context and spatial context when it comes to country information. On the one hand, country can be considered spatial information. On the other hand, it is coarse-grained and most works exploiting country information argue for a cultural perspective. Therefore, we discuss corresponding work here.

In an early work, \cite{mesnage_etal:womrad:2011} propose a social music recommender that leverages the music preferences of \textit{Facebook friendship connections} to make recommendations for the target user. 
The approach is based on the diffusion of songs that are given a high rating by a user through this user's friendship network. Diffusion stops when the song reaches a user who gives it a low rating.
For evaluation, the authors extract a list of one million tracks from Last.fm, and match those to YouTube videos. They implement their social recommendation approach in a Facebook app and compare its performance to that of a non-social recommender which recommends 
highly rated songs by users not in the target user's friendship network. In a study involving 68 users of their app, Mesnage et al.~show that 47\% of songs recommended by the social approach receive three or more stars, whereas this number is only 33\% for the non-social approach. Their social approach is limited, however, to users belonging to a (reasonably sized) social network. 
Users' relationships in a social network are also leveraged by \cite{DBLP:conf/dcai/Sanchez-MorenoP18}, who compute measures of \textit{social influence of users} from their friendship connections. These measures are integrated into a memory-based collaborative filtering algorithm to give more weights to influential users.  
S\'anchez-Moreno et al.~define the social influence of a user in two ways: their absolute number of friendship connections and the logarithm of this number. The resulting influence score of each user is integrated into a standard \textit{user-based CF} algorithm, to weight the similarity between the active user and their nearest neighbors. 
The resulting two approaches, integrating the two measures of social influence, are evaluated on the hetrec2011-lastfm-2k dataset~\cite{Cantador:RecSys2011},\footnote{The dataset is available at \url{http://ir.ii.uam.es/hetrec2011/datasets.html}.} containing artist listening information and friendship relations for about 12K Last.fm users.
The authors convert the artist playcounts given in the Last.fm data to ratings, by binning them for each user according to percentiles of the number of artists the user listened to.
They further investigate four similarity metrics to compute rating-based user similarity, i.e., cosine, Jaccard, Chebyshev, and Euclidean. In terms of normalized RMSE, their experiments show a substantially lower error for both CF approaches that use social influence in comparison to standard user-based CF, regardless of the adopted user similarity metric.
Limitations of this work include the relatively small dataset (less than 13K friendship relations, each user listened to only 9 artists on average). The authors furthermore provide no discussion of the suspiciously low normalized RMSE values (very close to zero) for some of their algorithmic variants, nor on the concrete acquisition procedure for the used data set (e.g., how users were sampled), which might distort results.

Another approach to integrate social information into an MRS is taken by \cite{DBLP:journals/mta/ChenYZ19} who propose a graph-based method incorporating social influence between users of a social network.
To this end, the authors first model a \textit{heterogeneous network} including entities such as songs, users, tags, or playlists. They compute several topological features for this network, which estimate indirect social influence between pairs of users.\footnote{As opposed to direct (or explicit) social influence such as a friendship connection, indirect social influence refers to a relationship between users by other means, e.g., liking the playlist of another user or attaching the same specific tag on a song.}
These measures include unweighted and weighted path count as well as random walks. 
They are integrated into a \textit{factor graphic model}, which is learned via gradient descent, and used to predict whether there exists a link between user and song nodes.
For evaluation, Chen et al.~crawl a dataset from Last.fm and show that the factor graphical model using weighted path count as distance measure between users outperforms the use of other distance measures as well as other baseline approaches, in terms of precision and recall. A severe limitation of this work is that important details on the origin of additional user features the authors claim to use (e.g., personality, friendship strength, and spatial features) is missing.

Another branch of research defines the social context of users through their cultural embedding in society, where the latter is often approximated by country of origin or residence.
For instance, \cite{zangerle:tismir:2020} propose a \textit{music-cultural user model}, based on which they approach the MRS task as a relevance classification problem, predicting whether the target user will like each song in a set of candidates for recommendation.
Each user is described by a list of features reflecting their individual music preferences (via acoustic features from Spotify\footnote{\url{https://developer.spotify.com/web-api/get-several-audio-features}}), cultural background (via Hofstede's cultural dimensions\footnote{\url{https://hi.hofstede-insights.com/national-culture}}), and socio-economic aspects (using the World Happiness Report\footnote{\url{https://worldhappiness.report}}). All of the target user's features are concatenated and fed into a relevance classifier to create recommendations. As classifier, Zangerle et al.~use XGBoost, a scalable tree boosting method.
Evaluation is performed on the standardized LFM-1b dataset of Last.fm users' listening records and demographics~\cite{10.1145/2911996.2912004},\footnote{The dataset is available at \url{http://www.cp.jku.at/datasets/LFM-1b}.} including more than 1B listening records of about 120K users. The combination of music preference, cultural, and socio-economic features outperforms using only parts of the user model as well as several baselines, both for  rating prediction (in terms of RMSE) and ranking (precision and recall). A limitation of the approach is the quite simple music preference model which is defined as the mean over the Spotify audio features of the user's listened tracks.

In another recent work, \cite{10.3389/frai.2020.508725} use a \textit{variational autoencoder} (VAE) architecture where country information is fed into the network through a gating mechanism to create culture-aware music recommendations. For this purpose, the authors first identify clusters of countries from their users' song listening behavior (playcount vectors over all songs), yielding archetypes of music-cultural regions. Based on these, four models with different representations of a user's cultural context are proposed: \textit{(1)} the user's country identifier, \textit{(2)} the user's cluster identifier, \textit{(3)} the distance between the user's music listening playcount vector to each cluster centroid, and \textit{(4)} the same distance but to each country centroid. Integrated into the VAE, Schedl et al.~evaluate the four models on the LFM-1b dataset of Last.fm listening records. They find that all four models significantly outperform a VAE model without context information as well as a most popular item recommender and implicit matrix factorization (with respect to precision, recall, and NDCG). However, no significant differences in performance between the four variants could be made out.

\subsection{Recommending sequences of music}\label{sec:sequences}

\begin{table*}[!tb]
\centering
\caption{\label{tab:overview_sequences}Overview of research works on \textit{recommending sequences of music}.}
\begin{tabular}{l | c | c | c | c | c | c}
\toprule
\multicolumn{1}{l}{\textbf{\footnotesize{Sub-goal/major method}}}  &\multicolumn{5}{c}{\textbf{\footnotesize{Level of Content}}} \\  \cline{2-6}
\multicolumn{1}{l}{\textbf{\footnotesize{}}}  &\multicolumn{1}{c}{\textbf{\footnotesize{Audio}}} &\multicolumn{1}{c}{\textbf{\footnotesize{EMD}}} &\multicolumn{1}{c}{\textbf{\footnotesize{EGC}}} &\multicolumn{1}{c}{\textbf{\footnotesize{UGC}}} &\multicolumn{1}{c}{\textbf{\footnotesize{DC}}} \\ \hline

\footnotesize{\textbullet \ Location and path-aware APG~\cite{miller2010geoshuffle}} & \cmark & \cmark &  \cmark & \cmark & \\
\footnotesize{\textbullet \ Markov chain modeling for APG~\cite{chen_etal:kdd:2012}} &\cmark & & \cmark & \cmark & & \\
\footnotesize{\textbullet \ Two-staged coherence-aware APG~\cite{Jannach:2015:BHH:2792838.2800182}} & & \cmark & & \cmark & \\
\footnotesize{\textbullet \ Long- and short-term preference-aware APG~\cite{DBLP:conf/um/KamehkhoshJL16}} & & & & \cmark & & \\
\footnotesize{\textbullet \ Attentive RNN leveraging tags ~\cite{DBLP:conf/recsys/SachdevaGP18}} & & & & \cmark & \\
\footnotesize{\textbullet \ Online learning to rank ~\cite{DBLP:conf/recsys/PereiraUPSZ19}} & \cmark & \cmark & & \cmark & \\
\footnotesize{\textbullet \ \changed{Efficient online learning to rank~\cite{DBLP:conf/www/ChavesPS22}}} & \cmark & \cmark & \cmark & & \\

\hline
\end{tabular}
\end{table*}

Often, users of MRSs are not only interested in receiving arbitrarily ordered music recommendations that do not constitute a certain meaningful sequence other than being ranked according to some prediction score. Instead, a common use case is a user's request to obtain a list of music recommendations that follow some semantics in their sequential ordering, e.g., transitional coherence~\cite{Schedl2018,Bonnin:2014:AGM:2658850.2652481}. This task of commonly referred to automatic playlist generation (APG). Content-driven approaches play a crucial role to accomplish this goal since they have access to and incorporate semantic descriptors that reflect how users perceive music, which is important to create a meaningful track sequence (e.g., according to tempo, rhythm, or musical style). The survey by Quadrana et al.~\cite{quadrana2018sequence} provides an interesting review of sequence-based approaches in various recommendation domains, including music.

{
An early work to APG addresses the task of context-aware playlist generation~\citep{miller2010geoshuffle}.
Unlike other media (such as videos), music consumption can happen in hand-busy or eye-busy situations such as while driving or concentrating on labor. People may listen to certain types of music while taking a specific routine path, such as cycling or driving to work. Miller et al.~\cite{miller2010geoshuffle} propose \textit{GeoShuffle}, an MRS that considers the impact of the locations where the user listens to music in her daily life. Their proposed system is a mobile-based MRS that takes as input the user's 
GPS data, where each location possesses its distinctive atmosphere or semantics. GeoShuffle augments the user's listening preferences with location-based and time-based information and generates a playlist based on the user's location, path, and historical choices.

Chen et al.~\cite{chen_etal:kdd:2012} model playlists as Markov chains (MCs) and propose logistic Markov Embeddings (LMEs) to learn the representations of songs for playlist prediction. In their method, the learning procedure exploits the weak ordering between tracks in playlists. The tracks are projected into a space such that the transition probabilities between tracks in a first-order MC are proportional to Euclidean distances. The resulting space can be exploited to generate new sequences (playlists) or continuing exiting ones. In \citep{moore2012learning}, the same
authors extend their approach by leveraging tag information. To account for the locality in track transitions, Chen et al.~\cite{chen2013multi} enhance LMEs by clustering items and adding cluster-level embeddings. Wu et al.~\cite{wu2013personalized} propose a personalized version of LME where the strength of the projection into the Euclidean space is affected by the strength of their relationship between user and item latent embeddings.
A number of similar works that exploit MCs for playlist generation include~\cite{he2009web,hosseinzadeh2015adapting,rendle2010factorizing,li2019music,eskandanian2018detecting}. 

Jannach et al.~\cite{Jannach:2015:BHH:2792838.2800182} propose an APG method that recommends music items that not only appeal to the general taste of the listener, but also are coherent with the~\textit{most recently} played tracks. 
The focus of the authors  
is on next-track recommendation, given a history of recent tracks. The proposed method is two-staged: first, a set of candidate music tracks that match the most recent items in the user's listening history are selected by a multi-faceted scoring method that takes into consideration attributes such as track co-occurrences, musical and metadata characteristics. Second, the top items of the candidate list are re-ranked using an optimization approach that attempts to minimize the difference between the recently listened tracks and the continuation in terms of one or more desired quality dimensions.

Kamehkhosh et al.\cite{DBLP:conf/um/KamehkhoshJL16} suggest a model for APG that distinguishes between the user's short-term and long-term preferences, where the latter is obtained from their social network. The authors compare and contrast several approaches to exploit multi-dimensional user-specific preference signals. This is accomplished by looking at songs that the user has previously consumed (track repetition), songs by artists the user has previously enjoyed (favorite artists), tracks that are semantically comparable to previously liked tracks (topic similarity), and tracks that are often played in conjunction with the user's favorite songs (track co-occurrence). Their approach also considers tracks and artists enjoyed by the user's friends on social networks. For the short-term preference analysis, a comparison between different playlist generation algorithms shows that an item-based KNN approach outperforms more complex algorithms such as BPR. To assess the importance of long-term preferences, the authors investigate the listening histories of 2,000 Last.fm and Twitter users, and analyze repetitions. Their study reveals that 
more than a quarter of users' listening events are repeated songs. 
Thus, for next-track selection, short-term preferences should govern the recommendation process. 
Notwithstanding, long-term preferences should be used to improve personalization quality.

{Also, recurrent neural networks (RNNs) have shown promising capacity to uncover sequential item relationships and to incorporate rich data
~\cite{quadrana2017personalizing}. Therefore, RNNs have naturally received attention in the MRS community, too. Sachdeva et al.~\cite{DBLP:conf/recsys/SachdevaGP18} incorporate song metadata in an attentive RNN
to better learn the user's short-term preferences and recommend next songs. In particular, an \textit{attention network} with  
bidirectional gated recurrent units (Bi-GRUs) is used with \textit{song tags} to detect semantic changes in the session.
User listening histories are uses for evaluation, with each element containing the user id, song name, artist name, and time stamp. The tags for each song were collected using the Last.fm API. The evaluation results reveal that \textit{(1)} attentive neural networks outperform several baselines, and \textit{(2)} approaches leveraging tags outperform those that do not by a large margin. This demonstrates the merits of semantic tags in representing short-term user preferences for sequential recommendation.}

{Recently, multi-armed bandits (MABs) have emerged as a technique to approach the sequential decision making problem. 
Pereira et al.~\cite{DBLP:conf/recsys/PereiraUPSZ19} propose an online learning to rank (L2R) system named counterfactual dueling bandits (CDBs) for sequential music recommendation. Their key insight is that the item space of music catalogs in modern streaming services is vast, hindering the ability of MRSs to provide personalized recommendations beyond the user's feedback. To address this issue, the proposed system encodes each object (user, song, and artist) in a low-dimensional feature space composed of collaborative and content-based signals, i.e., textual, social, and audio representations. On this space, the authors model dueling ranking models defined as arms. 

To determine the winning model, they leverage implicit feedback signals (plays and skips) to update the sequence of song recommendations dynamically. The validation of the system is performed on a Last.fm dataset and shows the merits of the approach with respect to various baselines from both effectiveness and efficiency perspectives. 
Authors conclude that 
CDBs learns a more effective recommendation by requiring less interaction data than other approaches.} 

\changed{A similar line of research is presented by \cite{DBLP:conf/www/ChavesPS22}, in which Chaves et al. propose a novel online learning-to-rank approach that efficiently explores the space of candidate recommendation models by limiting itself to the orthogonal complement of the subspace of prior underperforming exploration orientations. Music metadata (such as song title, artist name, and social tags) from MusicBrainz and audio characteristics (such as instrumentals, loudness, and intensity) from Spotify are considered as content-based features. At the same time, CF information  uses low-rank matrix factorization assuming binary user feedback on music and artists. An in-depth examination with Last.fm listening session simulations shows significant gains over state-of-the-art methods in terms of early-stage performance and overall long-term convergence. A potential limitation of this study is its reliance on extensive sampling to regulate exploration; an alternative would be to directly model distributions across rankings to obtain the same result.}

\subsection{Improving scalability and efficiency}\label{sec:scalability}

As a result of an ever increasing amount of multimedia music data it has become crucial to design MRS algorithms that are efficient to train and update, to make recommendations, and to deliver respective content, in the presence of both large item collections and number of users.
However, algorithmic scalability and efficiency of MRS are topics that are obviously more important from an industrial perspective than from an academic one. As a result, scientific publications on the topic are rather scarce compared to other challenges.

To evaluate the computational efficiency of a recommendation algorithm, the following measures can be used: \textit{(1)} training time of the model, \textit{(2)} prediction time for recommendations, and \textit{(3)} memory requirements~\cite{aggarwal2016evaluating,shani2011evaluating}.
Several strategies to improve efficiency of MRS algorithms have been developed, for instance, hashing of audio features~\citep{10.1145/1178677.1178699}, proactive caching of audio content~\citep{Koch:2017:PCM:3083187.3083197}, reducing the search space when computing similarities or identifying nearest neighbors~\citep{10.1145/1178677.1178699}, and incremental model training~\citep{DBLP:journals/taslp/YoshiiGKOO08}.

An early example based on \textit{hashing} can be found in \citep{Cai:2007:SMR:1291233.1291466}. The authors present a scalable MRS that follows a \textit{query-by-example} setting. 
The system uses an audio snippet of a seed song given by the user to search for the desired music in a catalog. Specifically, a music piece is first transformed to a \textit{music signature} sequence, where the signature represents the timbre of a local music segment. Afterward, to build a scalable indexing system, every signature is transformed into a hash code, using \textit{locality sensitive hashing}, where the parameters are adaptively estimated based on data scale. To create recommendations, the representative signatures from the snippets of a seed song are used to retrieve songs with matching melodies from the indexed dataset, therefore named recommendation by search. 
For relevance ranking, several criteria are considered, including the match ratio, temporal order, term weight, and matching confidence. The authors further present an application of their system for dynamic APG. Cai et al.~validate their system on music collections of varying sizes, ranging from 1K to 114K music pieces. They demonstrate its efficiency in terms of index size, memory requirements, and search time.
A limitation of this work is that it does not offer personalized recommendations.

Another approach to increase efficiency is \textit{proactive caching}, i.e.~prefetching content that the user is likely to consume next (listen in the case of MRS). Deciding which content to prefetch is challenging though, but the task is very similar to the recommendation problem, thus enabling the use of RS technology.
Koch et al.~\cite{Koch:2017:PCM:3083187.3083197} propose a prefetching approach for \textit{video streaming} websites, in particular YouTube, where more than 40\% of traffic is caused by watching music videos. To this end, content is prefetched based on, e.g., social user information or content properties, in a dedicated share of the cache while the remaining share is managed reactively~\cite{koch2017vfetch}. However, when it comes to music streaming, the task is more challenging than for videos. While it is easy to predict that a user will likely watch the next episode of a TV series on Netflix, for example, this does not hold in the case of music consumption. In fact, 
several studies have shown poor performance of approaches based on the user's subscription status and global item popularity~\cite{koch2017vfetch,wilk2016effectiveness,wilk2015potential}.

To approach this problem, Koch et al.~\cite{Koch:2017:PCM:3083187.3083197} propose a \textit{proactive caching pipeline for music videos}, which includes both CBF and CF recommenders to fill a share of the cache with suggested (music) videos.
For the CBF component, the authors extract all audio features available in the MIRtoolbox~\citep{10.1007/978-3-540-78246-9_31}, then drop highly correlated features and retain only those with low entropy. This results in around 300 features per video. 
Two approaches are then used to decide which items to prefetch: \textit{(1)} a personalized one that prefetches the most similar content to what the target used has watched and \textit{(2)} a user-independent one that prefetches items with the same genre or mood than the videos watched during the same hour of the day the request is made. The second approach adopts a genre and mood classifier trained on the audio features and Last.fm tags.
Also for the CF component two approaches are investigated. First, matrix factorization using \textit{alternating least squares} (ALS) is applied to implicit user feedback data (frequency of watching a video and percentage of the duration each video is watched). 
Second, a memory-based user-based CF approach, defining ratings as watch counts of videos, is adopted.  
Final recommendations are created by merging the recommendation lists produced by the different approaches and summing up the individual item scores. 
Koch et al.~evaluate their approach on a trace provided by a mobile network operator. After filtering videos that could not be matched to Last.fm artist and track names, or for which no Last.fm tags exist, they are left with about 14K YouTube videos watched by 5K users.

On this dataset, the authors find that the user-based CF recommender 
outperforms the other approaches, in particular the CBF ones, in terms of F1 score. 
On the other hand, they highlight that the user-independent CBF approach is more privacy-preserving.
A limitation of the work is the rather simple recommendation approaches, in particular the CBF-based ones.

Addressing the task of \textit{increasing efficiency of APG}, Knees et al.~\cite{10.1145/1178677.1178699} propose a method to accelerate similarity computation between songs, an important component of content-based APG systems. More precisely, their approach leverages artist similarities computed from web documents to restrict the search space for nearest neighbor search of tracks in an audio feature space. 
The addressed APG scenario is to create a \textit{circular playlist of all tracks} in a given collection, in such a way that subsequent tracks are as similar as possible. To this end, pairwise timbre-based audio similarity is first computed from all tracks' MFCCs.
A playlist is then created by adopting the minimum spanning tree heuristic for a \textit{traveling salesman problem} (TSP) defined by the audio-based track distance matrix.
To speed up estimating a solution for the TSP, Knees et al.~first retrieve the top web pages for each artist in the music collection, as returned by the Google search engine. These web pages are then represented as a single TF-IDF vector for each artist. The artist vectors are clustered using a self-organizing map (SOM), and only songs by artists belonging to the same or very similar clusters are considered candidates for the next track in the playlist. Formally, the track distance matrix is altered by setting to infinity the distances between all pairs of songs whose artists are found dissimilar according to the SOM clusters.
Evaluation is carried out on two collections of 3.5K and 2.5K tracks, respectively. The authors show that their approach for search space pruning not only reduces the number of necessary audio similarity computations drastically but the extent of this reduction can also be controlled by adapting the size of the used SOM.
A limitation of this work is the use of proprietary music collections for evaluation, which hinders reproducibility.

\begin{table*}[!tb]
\centering
\caption{\label{tab:overview_scalability}Overview of research works on \textit{improving scalability and efficiency}.}
\begin{tabular}{l | c | c | c | c | c | c}
\toprule
\multicolumn{1}{l}{\textbf{\footnotesize{Sub-goal/major method}}}  &\multicolumn{5}{c}{\textbf{\footnotesize{Level of Content}}} \\  \cline{2-6}
\multicolumn{1}{l}{\textbf{\footnotesize{}}}  &\multicolumn{1}{c}{\textbf{\footnotesize{Audio}}} &\multicolumn{1}{c}{\textbf{\footnotesize{EMD}}} &\multicolumn{1}{c}{\textbf{\footnotesize{EGC}}} &\multicolumn{1}{c}{\textbf{\footnotesize{UGC}}} &\multicolumn{1}{c}{\textbf{\footnotesize{DC}}} \\ \hline

\footnotesize{\textbullet \ Locality sensitive hashing of audio ~\cite{Cai:2007:SMR:1291233.1291466}} & \cmark & \cmark & & & & \\
\footnotesize{\textbullet \ Proactive caching of music videos ~\cite{Koch:2017:PCM:3083187.3083197}} & \cmark & & & \cmark & \cmark & \\
\footnotesize{\textbullet \ Accelerating audio similarity computation for APG~\cite{10.1145/1178677.1178699}} & \cmark & & \cmark & \cmark & & \\

\footnotesize{\textbullet \ Incremental adaptation of probabilistic generative model~\cite{DBLP:journals/taslp/YoshiiGKOO08}} & \cmark & \cmark & & & & \\

\hline
\end{tabular}
\end{table*}

Yoshi et al.~\cite{DBLP:journals/taslp/YoshiiGKOO08} approach the challenge of scalability in MRS by proposing a hybrid system to unify collaborative and content-based data. It relies on a \textit{probabilistic generative model}, supporting incremental model adaptation.
Yoshii et al.'s approach incrementally adapts a \textit{three-way aspect model}, which is an extension of probabilistic latent semantic analysis~\citep{DBLP:conf/uai/Hofmann99} and integrates audio as well as collaborative data.
To describe each track via audio features, its MFCCs are extracted and the parameters (mixture weights) of a fixed-size GMM learned from the entire collection are used as one input to the three-way aspect model.
Trained on these timbre-specific mixture weights and user ratings of each piece, the three-way aspect models learns latent variables, which the authors interpret as latent favorite genres of the users. For training, a variant of the expectation maximization (EM) algorithm is adopted.
Since the probabilistic model assumes statistical independence of all user profiles, the model can be updated in constant time for each newly added user ratings.
Evaluation is carried out on a collection of 358 Japanese chart songs. Ratings for these songs are obtained from Japanese Amazon from 316 users. 
Compared to model- and memory-based CF and CBF variants, the proposed generative model with incremental adaptation is very fast during update (10 minutes to train the base model versus 5 seconds to update it), while still achieving high accuracy.
Limitations of Yoshii et al.'s work include the sole use of MFCC features to describe audio and the small size and high bias of the music collection (only Japanese songs).

\subsection{Alleviating cold start}
\label{subsec:all_cs}
\begin{table*}[t!]
\centering
\caption{\label{tab:overview_coldstart}Overview of research works on \textit{alleviating cold start}.}
\begin{tabular}{l | c | c | c | c | c | c}
\toprule
\multicolumn{1}{l}{\textbf{\footnotesize{Sub-goal/major method}}}  &\multicolumn{5}{c}{\textbf{\footnotesize{Level of Content}}} \\  \cline{2-6}
\multicolumn{1}{l}{\textbf{\footnotesize{}}}  &\multicolumn{1}{c}{\textbf{\footnotesize{Audio}}} &\multicolumn{1}{c}{\textbf{\footnotesize{EMD}}} &\multicolumn{1}{c}{\textbf{\footnotesize{EGC}}} &\multicolumn{1}{c}{\textbf{\footnotesize{UGC}}} &\multicolumn{1}{c}{\textbf{\footnotesize{DC}}} \\ \hline
\footnotesize{\textbf{New item and user}} & & & & & & \\
\footnotesize{\textbullet \ Semantic enrichment using preference data (sparse rep.)~\cite{DBLP:conf/icmcs/SoleymaniAWV15}} & \cmark & & & & \\
\footnotesize{\textbullet \ Semantic enrichment using preference data (neural E2E)~\cite{DBLP:conf/icmcs/ChouYYJ17}} & \cmark & &\cmark & & \\
\footnotesize{\textbullet \ Semantic enrichment using tags (multi-modal PMF)~\cite{olivier_gouvert_2018_1492537}} & \cmark & & &\cmark & \\
\footnotesize{\textbullet \ \changed{SNNs trained on audio features to produce a query-by-example recommender~\cite{DBLP:conf/recsys/PulisB21}}} & \cmark & & & & \\
\footnotesize{\textbullet \ \changed{
Latent semantic analysis for cold-start playlist generation
~\cite{DBLP:journals/ijmir/YurekliKB21}}} & & \cmark  & \cmark & & \\
\footnotesize{\textbullet \ \changed{Metric
learning from audio and user embeddings with SNNs~\cite{DBLP:conf/icassp/0021LMG21}}} & \cmark & & & & \\

\footnotesize{\textbf{Data sparsity}} & & & & & & \\
\footnotesize{\textbullet \ Semantic enrichment using tags (three-way tensor)~\cite{DBLP:journals/taslp/NanopoulosRSM10}} & \cmark & & &\cmark & \\
\hline
\end{tabular}
\end{table*}

The cold start (CS) problem refers to situations where due to lack of sufficient user--item interactions, classical recommendation models based on CF cannot provide (useful) recommendations.
CS is reflected in different problems: \textit{data sparsity} (when the entire user--item matrix has a low amount of interactions), \textit{new user} (when a new user enters the system, having no or few interactions in her profile), or \textit{new item} 
(when new items are added to the catalog, lacking sufficient user interactions). A low popularity of certain categories of items can also cause similar problems, e.g., particular music styles or artists may attract much fewer interactions compared to others. 
While CS primarily concerns CF approaches, some works, such as~\cite{DBLP:journals/taslp/NanopoulosRSM10,DBLP:conf/icmcs/SoleymaniAWV15}, refer 
to the lack of social tags' availability, which can hinder content-driven MRS performance. 
Our onion model is a useful tool to explain this aspect. 
The farther we go away from the onion model's center, the higher are the chances that we experience CS. For instance, UGC (e.g., social tags) compared to EMD (e.g., performing artist) are more susceptible to CS, because when a piece of music is released it is accompanied by EMD from record labels, whereas it will take a while for UGC to be created. 
Similarly, considering EGC versus audio signal, the latter constitutes a perfect source to alleviate CS issues in content-based MRSs, because audio features can always be extracted from the raw audio, even for content that does not contain any descriptive tags yet.

\changed{Pulis et al.~\cite{DBLP:conf/recsys/PulisB21} propose using Siamese neural networks (SNNs) to determine the degree to which two audio clips are similar. To this end, the Mel spectrograms are fed into a SNN, 
which consists of two identical CNNs. After running each pair of tracks through a CNN, the results are compared to see how similar they are. These were trained to recognize musical similarities between tracks by using audio from the Free Music Archive, with the use of genre as a heuristic. As shown by the results, the developed model can accurately compute similarity for a large number of music genres with a small average loss; specifically, 81.64\% accuracy was achieved on a dataset of 6,400 music pairs. The suggested method offers recommendations based on audio content, despite the fact that even new artists do not have a sufficient number of listeners. In fact, as shown by the experimental validation, 55\% of the recommended songs had less than 1,500 plays, indicating that the proposed content-based method can provide fairer exposure to all artists based on their music, regardless of their popularity. To better serve users, a query-by-multiple-example (QBME) music recommendation system was built and trained with the proposed content-based similarity metric.}

\changed{To alleviate the cold-start issue in APC, Y{\"{u}}rekli et al. in~\cite{DBLP:journals/ijmir/YurekliKB21} provide an ad hoc retrieval technique with the following main idea: given a query (playlist title), find nearby clusters with the LSA model, and weight/rank tracks for inclusion in the playlist. In particular, the proposed method is comprised of four stages: (i) textual playlist clustering, (ii) LSA modeling, (iii) nearby cluster retrieval, and (iv) track weighting. The system initially collects playlists based on textual groupings of user-generated titles. An LSA model is developed to evaluate pairwise document similarity by describing clusters as documents and phrases. Given a cold-start playlist (i.e., only a playlist title is available), the algorithm extracts relevant documents that match the query terms.
Once the surrounding clusters have been identified through neighborhood selection, candidate tracks within these clusters are rated using track weighting schemes to deliver final recommendations to the target playlist. For experimental validation, the authors conduct multiple experiments on the MPD, which contains one million real-world Spotify music playlists, and demonstrate promising accuracy on large test sets.}

\changed{Finally in~\cite{DBLP:conf/icassp/0021LMG21}, Chen et al. offer a methodology for learning audio embeddings that can be used to address the cold-start problem in MRSs by representing recently released tracks. In order to construct an embedding that represents a user's musical preferences, they initially study the user's listening history and demographics. Using metric learning with Siamese networks, the user embedding, and audio data from the user's favorite and least favorite tracks, an audio embedding for each track may be obtained. Specifically, log-mel spectrograms of audio inputs are fed into a SNN consisting of CNNs with shared weights and setups. By calculating the similarity between the track's audio embedding and multiple user embeddings, the authors identify the ideal set of users to recommend for a new track. As for the limitations, the generalization capability of the learnt audio embedding in various MIR tasks can be investigated further. To iteratively enhance the proposed model, one can make full use of offline metrics. }

\setlength{\parindent}{15pt} \textbf{New item and user:} {Soleymani et al.~\cite{DBLP:conf/icmcs/SoleymaniAWV15} show how audio CB features and sparse modeling can be combined to learn new types of attributes. These attributes can then be automatically estimated for rating prediction of cold items. The authors demonstrate the superiority of their approach for cold item recommendation with respect to several baselines (user's average rating, genre-based, artist-based, or pure CBF), using RMSE as a performance metric.}

{Some approaches have tried to address the CS problem using deep neural models. For instance, Chou et al.~\cite{DBLP:conf/icmcs/ChouYYJ17} propose an E2E deep neural model to cope with both new user and item problem under the same framework. The proposed system is named conditional preference network (CPN), which  leverages three information sources: \textit{(1)} the available interactions, \textit{(2)} the result of an introductory survey used to deal with the new user problem, \textit{(3)} content feature of the items. CPN incorporates this information while training a neural model to address both new user/item problem simultaneously. As for the introductory survey, the authors rely on a previous offline framework by Rashid et al.~\cite{DBLP:journals/sigkdd/RashidKR08} that is based on the assumption that new users will always pick few items with the highest play counts. 
Validation of the system is carried out on the MSD dataset using accuracy metrics.}

{Gouvert et al.~\cite{olivier_gouvert_2018_1492537} present a matrix co-factorization method based on PMF, which jointly addresses two tasks: \textit{(1)} song recommendation and \textit{(2)} tag labeling. The system is named multi-modal PMF and can be used for cold item recommendations (without interactions), by leveraging tags associated with the items. The proposed method learns the feature vectors of the cold and warm items by jointly optimizing a loss function defined based on interaction data and social tags. The validation is done on two datasets, MSD and a proprietary dataset from Last.fm. The authors show that their proposed approach is not prone to CS issues since it can swiftly switch to another modality when the core modality is not available.}

\setlength{\parindent}{15pt} \textbf{Data sparsity:}
 Similarly, Nanopoulos et al.~\cite{DBLP:journals/taslp/NanopoulosRSM10} propose three-order tensors to capture three-way correlations between users, items, and tags. The authors show the effectiveness of their approach with social tagging data from Last.fm for dealing with data sparsity, here referred to as the lack of availability of social tags. 
It is also shown that combining social tags with acoustic features improves recommendation quality compared to the sole use of each of these. \changed{Several recent publications \cite{sanchez2021dynamic,sanchez2020using,yang2018social,xu2017tag}discuss the use of social tags to characterize music. In some of these articles, the tags are processed using word embedding techniques. A significant disadvantage of the aforementioned studies is that they frequently do not utilize user-item interaction data.}

\section{Conclusions and Open Grand Challenges}\label{sec:conclusion}

In this survey, we reviewed a total of \changed{55} articles on content-driven music recommendation and categorized them according to our proposed onion model of music content data. The model structures available sources of content and aims to provide a better understanding of them in the context of music recommendation. More precisely, the proposed model is a hierarchical model whose  
layers represent content categories, starting with the audio signal at its core and gradually adding layers that exhibit higher semantic valence and higher subjectivity:
signal $\rightarrow$ embedded metadata (EMD) $\rightarrow$ expert-generated content (EGC) $\rightarrow$  user-generated content (UGC) $\rightarrow$ derivative content (DC). 
Moving from the most inner layer 
towards the outer layers, we provide a characterization of each content category according to the contained features' temporal component, semantics, subjectivity, cultural context, community, cold start, data diversity, quality, and credibility. 
The main advantages of the proposed onion model, according to which we organized the reviewed literature are: \textit{(1)} it categorizes all kinds of musical data in a meaningful way and \textit{(2)} it highlights the trade-offs according to the characteristics described above. We deem this way of looking at music content data useful, and assume that it will stimulate further research on the topic of content-driven music recommendation.

From the reviewed literature, we identified six overarching challenges, according to which we organize our survey from a high-level perspective:  
increasing recommendation diversity and novelty, 
providing transparency and explanations, 
accomplishing context-awareness, 
recommending sequences of music, 
improving scalability and efficiency, and 
alleviating cold start.
Complementarily, we categorize and discuss the literature based on the layers of our unified onion model, based on its goals, and the main methodological choices.

Even though much research has been devoted to approach these challenges, there persist several grand challenges that still need to be addressed in future research endeavors. In the following, we present some of the most pressing ones, according to our experience and perspective.

First, it became obvious during the review of relevant articles for this survey that the outermost category in our onion model of music content, i.e., derivative content, has largely been neglected in the context of MRSs so far. Therefore, \textit{acquiring, processing, understanding, and integrating derivative content} into MRSs is one big open challenge. While the immediate utility of derivative content such as remixes, cover versions, or parodies of songs may barely be tangible for a MRS, considering such content within a larger socio-cultural perspective may reinforce research on culture-aware MRSs and beyond.

A second grand challenge is \emph{understanding the cognitive processes in human decision making} given a large set of choices, and --- strongly related to that --- how corresponding psychological models of cognition should be considered in RSs research. These questions connect to the recent research area of psychology-informed RS~\cite{lex2021psychology}. Music content information can support such psychology-informed models because processes of human memory are often tied to content categories (e.g., acoustic characteristics or music genre).

Another grand challenge of content-driven MRSs is the \textit{multi-faceted definition of diversity and novelty.} 
{In MRS research, there are different types of content features, according to the onion model, that play a key role in users' affinity towards the songs (e.g., rhythm, artist, genre, lyrics, reception of the general public).
This raised the need for defining multi-faceted diversity and novelty metrics in order to evaluate the recommendation quality under these complementary perspectives. 
While there exists a multitude of corresponding metrics covering different perspectives (content, interaction, intra- vs.~inter-list, etc.), their importance for real-world MRSs and how they translate to user satisfaction is still unclear.

Another major open challenges is \textit{mitigating societal biases and improving fairness}~\cite{deldjoo2021explaining,deldjoo2019recommender,Ekstrand2021fairness}.
{
Fairness can be regarded from different stakeholders' perspectives. In particular, the main stakeholder in MRSs are end users (consumers) and items (or, more precisely, content creators or providers)~\cite{deldjoo2023fairness}. Here, some features in the onion model can be seen as more sensitive than others when considering items. For example, by regarding gender as a sensitive attribute~\cite{ferraro2021break}, the gender of the artist recommended can be an essential aspect to study the fairness of MRSs, and how much MRSs provide equal gender opportunity. 
A similar argument can be made for other protected attributes of users or items. 
Investigating the trade-off between user fairness, item fairness, and accuracy of MRSs, and how we can maintain a balance between these factors, are interesting open research directions.

{Finally, an interesting emerging topic is 
\textit{conversational music recommender systems} (CMRSs). Unlike traditional MRSs, which commonly learn users' long-term music preferences from historical interactions, CMRSs can elicit preferences towards context, including emotion and geo-temporal aspects~\cite{zhou2018musicrobot}. How different content levels can be leveraged by CMRSs, how different interaction modes with the system (e.g., natural language vs.~spoken audio) should be considered, and how the recommendations are presented to the user (e.g., uni-modal vs.~multi-modal)~\cite{deldjoo2021towards,deldjoo2022multimediach} can play a crucial role in building effective music-oriented chatbots for MRSs, but largely remain open challenges.}

To wrap up, we hope and believe that the presented survey will serve as a standard for categorizing music content, as a reference to the evolution and state of the art in content-driven music recommendation, and as a stimulus for further research in this exciting domain.

\section*{Acknowledgments}
This work is supported by the Austrian Science Fund (FWF): P33526 and DFH-23; and by the State of Upper Austria and the Federal Ministry of Education, Science, and Research, through grant LIT-2020-9-SEE-113.

\bibliographystyle{elsarticle-num}
\bibliography{main}

\end{document}